\documentclass[12pt]{article}
\usepackage{epsfig} 
\usepackage{graphics}
\usepackage{a4wide}
\usepackage{enumerate}
\usepackage{amssymb,amsmath}

\newcommand{\cir}[1]{\raisebox{.6pt}{\textcircled{\raisebox{-1.1pt} {#1}}}}

\newcommand{\half}{{\scriptstyle\frac{1}{2}}}
\newcommand{\RR}{{\mathbb{R}}}
\newcommand{\CC}{{\mathbb{C}}}
\newcommand{\ZZ}{{\mathbb{Z}}}
\newcommand{\pa}{\partial}
\newcommand{\tr}{\mathop{\rm tr}\nolimits}

\newcommand{\vr}{{\bf r}}
\newcommand{\ve}{{\bf e}}

\newcommand{\vb}{{\bf b}}

\newcommand{\ii}{{\rm i}}
\newcommand{\ee}{{\rm e}}
\newcommand{\UU}{{\rm U}}
\newcommand{\st}{\tilde{s}}
\newcommand{\vep}{\varepsilon}

\title{Moduli of Monopole Walls and Amoebas}
\author{
Sergey A.\ Cherkis\\
\vspace{-8pt}
\it\small Department of Mathematics\\
\vspace{-8pt}
\it\small University of Arizona,\\
\vspace{-5pt}
\it\small Tucson AZ, 85721-0089, USA\\
\tt\small cherkis@math.arizona.edu
\and
Richard\ S.\ Ward\\
\vspace{-8pt}
\it\small Department of Mathematical Sciences\\
\vspace{-8pt}
\it\small University of Durham,\\
\vspace{-5pt}
\it\small Durham DH1 3LE, UK\\
\tt\small richard.ward@durham.ac.uk
}
\date{}

\begin{document}
\begin{titlepage}

\renewcommand{\thepage}{ }

\maketitle
\abstract{We study doubly-periodic monopoles, also called monopole walls, determining their spectral data and computing the dimensions of their moduli spaces.  Using spectral data we identify the moduli, and compare our results with a perturbative analysis.  We also identify an $SL(2,\mathbb{Z})$ action on monopole walls, in which the $S$ transformation corresponds to the Nahm transform.}
\end{titlepage}

\tableofcontents
\section{Introduction and Motivation}
The classical dynamics of monopoles was found to be intimately related to quantum gauge
theories in three \cite{Seiberg:1996nz, Chalmers:1996xh, Hanany:1996ie} and four
\cite{Cherkis:2000ft} dimensions.  For example, the moduli space of $n$
Bogomolny-Prasad-Sommerfield (BPS) monopoles with the gauge group $SU(2)$ is identified
with the moduli space of vacua of the three-dimensional $SU(n)$ Yang-Mills theory with
eight  supercharges \cite{Seiberg:1996nz}.  The moduli space of $n$ periodic $SU(2)$
BPS monopoles, on the other hand, is isometric to the space of vacua of the $SU(n)$
Seiberg-Witten theory on $\mathbb{R}^3\times S^1.$  In this paper we pursue this line
of thought, and explore BPS monopoles with two periodic directions.  Such monopoles
are also referred to as doubly-periodic monopoles or as monopole walls.
We use these two names interchangeably in this paper.

Monopole walls may be viewed as domain walls separating two
constant magnetic field phases, and in that context are linked to monopole bags \cite{Bol06}, which have been the subject of several recent studies \cite{LW09, Har11, Sut11, Man11, EG11}.
The walls which occur as the surface of monopole bags tend to have
(approximate) hexagonal symmetry, whereas in this paper we use
square symmetry, for simplicity. However, our analysis should 
extend to the hexagonal case.

Doubly-periodic monopoles are also related to quantum gauge theories via a chain
of string theory dualities.  Before we outline these dualities in
Section~\ref{Dualities}, let us define the doubly-periodic monopole problem
we consider.

\subsection{Monopole Wall}
A doubly-periodic BPS monopole is a hermitian
bundle $E\rightarrow T^2\times\mathbb{R}$ with a connection one-form $A$ and an
endomorphism $\Phi$ called the Higgs field.  The pair $(A, \Phi)$ satisfies the
Bogomolny equation
\begin{equation}\label{Bogomolny}
   *D_A\Phi=-F,
\end{equation}
where the covariant differential  $D_A$ is defined by $D_A\Phi=d\Phi+[A,\Phi]$,
and the curvature of the connection is $F=dA+A\wedge A$. We introduce affine
coordinates $x^1=x$ and $x^2=y$ on the torus $T^2=\mathbb{R}^2/\mathbb{Z}^2$,
each having period 1, and a coordinate $x^3=z$ on $\mathbb{R}.$  A priori, the
gauge group is U($n$), so that in any given trivialization $A$ is a one-form
and $\Phi$ is a function on an open chart of 
$T^2\times\mathbb{R}$, each valued in antihermitian $n\times n$ matrices.
For U(1) monopoles, rather than working with pure-imaginary fields, we let
$\Phi= \ii \phi$ and $A=\ii a=\ii(a_x dx+a_y dy+a_z dz)$, so that $\phi$ and $a$
are real-valued.  The energy density of a monopole wall is 
${\cal E}=-\half\tr\left[(D_j\Phi)^2+(B_j)^2\right]=-\frac{1}{2}\Delta |\Phi|^2$.

We also discuss SU($n$) monopole walls, so that $A$ and $\Phi$ are traceless,
and $E$ is a vector bundle with SU($n$) structure group.
Note that the tracefree part of a U($n$)
monopole defines a monopole with gauge group $\UU(n)/\UU(1)={\rm SU}(n)/\ZZ_n$,
which is not the same as an SU($n$) monopole \cite{Cherkis:2000ft}.
For example, the tracefree part of a U(2) monopole is an SO(3) monopole,
and in the periodic case this may or may not be an SU(2) monopole.

\subsubsection{Asymptotic Conditions}\label{Sec:ABC}
It is important to specify the boundary conditions as $|z|\rightarrow\infty$.
Before we do so, let us consider some simple abelian solutions of the Bogomolny equation.
If $E$ is a line bundle, then the gauge group is U(1) and Eq.~\eqref{Bogomolny}
is linear.  It implies that $\Phi$ is a harmonic function on $T^2\times\mathbb{R}.$ 
One such possible function is linear, leading to a {\em constant-energy solution}:
\begin{align}\label{ConstE}
  \phi&=2\pi  (Q z+M),& a&=2\pi  (Q y \,dx-p\,dx-q\,dy).
\end{align}
For $A$ to be a connection, $Q$ has to be an integer equal to the Chern class of
the bundle on the torus. The parameters $(M,p,q)$ are real constants, with $p,q\in[0,1)$.

As the abelian problem is linear, we can expand in Fourier modes along the periodic
directions. A nonzero Fourier mode is labelled by two integers $m_1$ and $m_2$,
and has the form
\begin{equation}\label{Mode}
\begin{split}
\phi&=  \sin (2\pi m_1 x) \sin( 2\pi m_2 y) \exp(2\pi m_{12} z),\\ 
a&= \frac{\exp(2\pi m_{12} z)}{m_{12}}
\left[m_2 \sin( 2\pi m_1 x)\cos( 2\pi m_2 y)  dx
   - m_1 \cos( 2\pi m_1 x)\sin( 2\pi m_2 y)  dy\right],
\end{split}
\end{equation}
where $m_{12}=\sqrt{m_1^2+m_2^2}$;
or a similar form with $\cos$ replacing some $\sin$ functions and vice versa.

While the former solution \eqref{ConstE}  has constant energy density, the latter
solution \eqref{Mode} has energy density
${\cal E}=4\pi^2\left(m_1^2\sin^2(2\pi m_2 y)+m_2^2\sin(2\pi x)\right)
  \exp(4\pi m_{12} z)$ growing exponentially at infinity.
In order to have some control over the solutions and their moduli, we model our
asymptotic conditions on the constant-energy solution, permitting at most linear
growth of the Higgs field at infinity.

We assume that asymptotically the eigenvalues of the U($n$) Higgs field $\Phi$ are
\begin{equation}\label{BC}
{\rm EigVal}\ \Phi=\left\{2\pi\ii\left(Q_{\pm,l}z+M_{\pm,l}\right)
                         +o(1/z)\,|\,l=1,\ldots,n\right\},
\end{equation}
where $Q_{\pm,l}$ and $M_{\pm,l}$ are real constants.  We call $Q_{\pm,l}$
monopole-wall charges; in fact, they are rational numbers.  Given \eqref{BC} for
large $|z|$, the vector bundle $E$ splits into eigenbundles of $\Phi.$ If there
are $ f_\pm$ distinct charge values $Q_{\pm j}$ as $z\rightarrow\pm\infty,$  
with $j=1,\ldots, f_{\pm}$; then as $z\rightarrow\pm\infty$
we have $E|_z=\mathop{\oplus}_{j=1}^{ f_\pm} E_{\pm j}$; and the Chern
number of $E_{\pm j}$ is
\begin{equation}
\int_{T_z} c_1(E_{\pm j})=\frac{\ii}{2\pi}\int_{T_z} {\rm tr}\, F_{\pm j}=
   -\frac{\ii}{2\pi}\int_{T_z} {\rm tr}\, *D\Phi_{\pm j}=
     {\rm rk}(E_{\pm j})\, Q_{\pm j}.
\end{equation}
Since the Chern number is an integer, $Q_{\pm j}$ has to be rational, with
${\rm rk}(E_{\pm j})$ divisible by the denominator of $Q_{\pm j}$:
\begin{align}
Q_{\pm j}&=\frac{\alpha_{\pm j}}{\beta_{\pm j}},&
{\rm rk}(E_{\pm j})&=r_{\pm j} \beta_{\pm j}.
\end{align}
Thus in the set $\{ Q_{+,l}\,|\, l=1,\ldots, n\}$ a given value $Q_{+j}$ appears
$r_{+j}$ times, and analogously the set $\{ Q_{-,l}\,|\, l=1,\ldots, n\}$ contains
a value $Q_{-j}$ exactly $r_{-j}$ times.

Including the subleading behaviour of the eigenvalues of $\Phi$ allows one to
potentially further split $E_{\pm e}$ into $r_{\pm e}$ subbundles
$E_{\pm j}=E_{\pm j}^1\oplus E_{\pm j}^2\oplus\ldots\oplus 
   E_{\pm j}^{r_{\pm j}}$,
each with Chern number $c_1(E_{\pm j}^{\nu})=\alpha_{\pm j}$, and with the
corresponding eigenvalue of the Higgs field $\Phi$ being
$2\pi \ii\left(Q_{\pm j}z+M_{\pm j}^\nu\right)+o(1/z)$ with
$\nu=1,2,\ldots, r_{\pm j}$.  Some of the values $M_{\pm j}^\nu$ can coincide,
but generically they are distinct.

In addition to the behaviour of the Higgs field $\Phi$, we also fix the eigenvalues
$\ee^{2\pi i p_{\pm,l}}$ of the asymptotic holonomy around the $x$-direction
at $y=0$, in a gauge in which the components of $A$ are $x$-periodic; and the eigenvalues  
$\ee^{2\pi i q_{\pm,l}}$ of the asymptotic holonomy around the $y$-direction
at $x=0$, in a gauge in which the components of $A$ are $y$-periodic.  Similarly to
the way of labeling $M$ as $M_{\pm,l}$ with $l=1,\ldots,n$, or labelling the same values as $M_{\pm j}^\nu$
with  $j=1,\ldots,\pm f$ and $\nu=1,\ldots, r_{\pm j}$, we label the holonomy
parameters as $p_{\pm,l}$ and $q_{\pm,l}$ or as $p_{\pm j}^\nu$ and
$q_{\pm j}^\nu$ with  $j=1,\ldots, f_\pm$ and $\nu=1,\ldots, r_{\pm e}$.  (Notice the subscript comma signifying the difference in labelling.)

\subsubsection{Singularities}\label{Sec:Sing}
One might limit the scope to considering monopoles with the above boundary conditions
that are completely smooth in the interior.  Here, however, we would like to allow
for Dirac-type singularities; this allows us to have a wider variety of interesting
moduli spaces, and leads to a complete picture of the Nahm transform.

To begin with, let us consider an example of a basic Dirac monopole wall. Its charges
are $Q_{-}=0$ and $Q_{+}=1$, and the fields are
\begin{eqnarray}
  \phi &=& \phi_0+\pi z-\frac{1}{2r}+\frac{1}{2}\sum_{j,k\in\ZZ}
          \left[\frac{1}{e_{jk}}-\frac{1}{r_{jk}}\right],\label{dirac_phi1}\\
a_{+} &=& \frac{1}{2}\sum_{j,k\in\ZZ}\frac{(y-k)dx+(j-x)dy}{r_{jk}(z+r_{jk})}
        +\frac{\pi}{2}(3y\,dx+x\,dy)\ \mbox{\ for $z\geq0$},\label{dirac_gaugepotplus} \\
a_{-} &=& \frac{1}{2}\sum_{j,k\in\ZZ}\frac{(y-k)dx+(j-x)dy}{r_{jk}(z-r_{jk})}
        +\frac{\pi}{2}(y\,dx-x\,dy)\ \mbox{\ for $z<0$}.\label{dirac_gaugepotminus}
\end{eqnarray}
Here $\phi_0$ is a constant, $\vr=(x,y,z)$ and $r=|\vr|$,
$\ve_{jk}=(j,k,0)$ and $e_{jk}=|\ve_{jk}|$, $r_{jk}=|\vr-\ve_{jk}|$,
and the $j=k=0$ term is excluded from the double sum in (\ref{dirac_phi1}).
The extra linear terms are chosen to ensure that the field behaves
like the constant-energy field (\ref{ConstE}) as $z\to\pm\infty$.
The gauge potentials $a_{+}$ and $a_{-}$ are related by a
gauge transformation (singular at $x=y=0$) across $z=0$. 
The series in Eq.~\eqref{dirac_phi1} gives a much studied doubly-periodic Green's function.
It is converging very slowly, however, a number of fast converging representations for it can 
be found in the literature.  See for example \cite{Linton} for exponentially fast converging representations of $\phi$ and for the value of $\phi_0$ that ensures that $\phi\rightarrow0$ as $z\rightarrow -\infty.$

Of course one can superimpose a number of such walls, for example a 2-pole wall
with negative singularities at $\vr=\vr_{-,1}$ and $\vr=\vr_{-,2}$, having
$Q_{-}=-1$ and $Q_+=+1$, and $\phi$ of the form
\begin{equation} \label{dirac_phi2}
\phi=\phi_0-\frac{1}{2|\vr-\vr_{-,1}|}-\frac{1}{2|\vr-\vr_{-,2}|}
    +\frac{1}{2}\sum_{j,k\in\ZZ}
         \left[\frac{2}{e_{jk}}-\frac{1}{|\vr-\vr_{-,1}-\ve_{jk}|}
                 -\frac{1}{|\vr-\vr_{-,2}-\ve_{jk}|}\right],
\end{equation}
with analogous expressions for the gauge potential.

For a general U($n$) monopole wall, we allow prescribed Dirac singularities.
At some predetermined positions $\vr_{+,\nu}$ for positive and
$\vr_{-,\nu}$ for negative singularities,  we permit the Higgs field
to diverge respectively as
\begin{align}
\Phi&=\ii \left(\begin{array}{cc}
\frac{+1}{2 |\vr-\vr_{+,\nu}|} & 0_{1\times(n-1)}  \\
 0_{(n-1)\times 1} & 0_{(n-1)\times(n-1)}
\end{array}\right)+O(|\vr-\vr_{+,\nu}|),\\
\Phi&=\ii \left(\begin{array}{cc}
\frac{-1}{2 |\vr-\vr_{-,\nu}|} &  0_{1\times(n-1)}  \\
 0_{(n-1)\times 1} & 0_{(n-1)\times(n-1)}
\end{array}\right)+O(|\vr-\vr_{-,\nu}|).
\end{align}

\subsection{Moduli and Parameters}\label{Sec:Moduli}
For any set of asymptotic data $(Q_{\pm}, M_{\pm}, p_{\pm},q_{\pm})$
and positions of Dirac singularities $\vr_{\pm}$, the space of all solutions
satisfying these conditions (whenever it is nonempty) forms a moduli space
${\cal M}$ or ${\cal M}_{Q,M,p,q,\vr}$ with a hyperk\"ahler metric.  In this paper,
we compute the dimension of ${\cal M}$ and introduce two sets of natural
coordinates on it.   We also establish an isometric action of the modular
group\footnote{We note here that this group is not the mapping class group of the torus of the base space $T^2\times\mathbb{R}.$} $SL(2,\mathbb{Z})$ on the set of all monopole-wall moduli spaces.

From the point of view of the moduli space ${\cal M}={\cal M}_{Q,M,p,q,\vr}$ itself,
the parameters $Q_{\pm}, M_{\pm}, p_{\pm}, q_{\pm}$ and $\vr_\pm$
appearing in the monopole boundary conditions determine its geometry:
for example the sizes of compact cycles and the asymptotic form of the metric.
We would like to distinguish {\em essential parameters} from
{\em superficial parameters}.  Variation of superficial parameters does not influence
the geometry of the moduli space ${\cal M}_{Q,M,p,q}$\,, while any variation
of the remaining, essential,  parameters does.  For example shifting the Higgs
field by a constant $C$ amounts to $M_\pm\mapsto M_\pm+C$; this does not change
the moduli space and is, therefore, superficial.  Another example is a translation
of the solution in the $z$-direction: it produces the change
$M_\pm\mapsto M_\pm+C Q_\pm$ and also is superficial. 

The asymptotic parameters are constrained by relations. A simple example is provided
by Dirac monopole walls, such as in
Eqs.~(\ref{dirac_phi1},\ref{dirac_gaugepotplus},\ref{dirac_gaugepotminus})
or Eq.~(\ref{dirac_phi2}), the examples of U(1) 1- and 2-pole monopole walls.
A U(1) $r_0$-pole monopole wall field depends on
$3 r_0+3$ parameters, of which $3r_0$ correspond to the location of the poles.
The remaining three are asymptotic data as $z\to\infty$: namely $(M_{+},p_{+},q_{+})$,
where $M_{+}$ is defined in (\ref{BC}), $p_{+}\in[0,1)$ corresponds to the holonomy
of the gauge field in the $x$-direction at $y=0$ as $z\to\infty$,
and $q_{+}\in[0,1)$ similarly
corresponds to the holonomy in the $y$-direction at $x=0.$ (The analogous parameters
$(M_{-},p_{-},q_{-})$ as $z\to-\infty$ are determined in terms of these $3r_0+3$
ones.) In general, one can have $r_{+0}\geq0$ positive poles and $r_{-0}\geq0$
negative poles, with $r_0=r_{+0} + r_{-0}$ being the total number of singularities.
The four integers $r_{\pm0}$, $Q_\pm$ are related by $Q_+ - Q_- = r_{-0} - r_{+0}$.
The three examples above -- the constant energy solution \eqref{ConstE} and 1- and 2-pole
Dirac walls (\ref{dirac_phi1},\ref{dirac_gaugepotplus},\ref{dirac_gaugepotminus})
and (\ref{dirac_phi2}) -- all have $r_{+0}=0$, and have $r_0=r_{-0}=0,1,2$ respectively.

Let us list all of the charges in the following manner: 
\begin{equation}
(Q_{,l})=(Q_{-,1},Q_{-,2},\ldots,Q_{-,n}, Q_{+,1}, Q_{+,2},\ldots, Q_{+,n}),
\end{equation} 
with $Q_{-,1}\geq Q_{-,2}\geq \ldots\geq Q_{-,n}$ and $ Q_{+,1}\geq Q_{+,2}\geq\ldots\geq Q_{+,n}$.  Here the index $l$ ranges from $1$ to $2n$. 
In the following we establish that the asymptotic and singularity conditions have to satisfy
\begin{align}\label{Closedness}
\sum_{j=1}^{f_-}r_{-j}\beta_{-j}&=\sum_{j=1}^{f_+}r_{+j}\beta_{+j}=n,&
r_{-0}+\sum_{j=1}^{f_-}r_{-j}\alpha_{-j}&=r_{+0}+\sum_{j=1}^{f_+}r_{+j}\alpha_{+j},
\end{align}
\begin{align}\label{Vieta}
\sum_{\nu=1}^{r_+} z_{+,\nu}-\sum_{\nu=1}^{r_-} z_{-,\nu}=\sum_{l=1}^n M_{+,l}-\sum_{l=1}^n M_{-,l}, \\
\sum_{\pm}\sum_{\nu=1}^{r_\pm}  \pm y_{\pm,\nu}+\sum_{\pm}\sum_{l=1}^n \pm p_{\pm,l}+\frac{1}{2}\sum_{\stackrel{l_1,l_2=1}
  {\mbox{\tiny $l_1<l_2$}}}^{2n} (Q_{,l_1}-Q_{,l_2})\in\mathbb{Z},\\
\sum_{\pm}\sum_{\nu=1}^{r_\pm}  \pm x_{\pm,\nu}+\sum_{\pm}\sum_{l=1}^n \pm q_{\pm,l}+\frac{1}{2}\sum_{\stackrel{l_1,l_2=1}
  {\mbox{\tiny $l_1<l_2$}}}^{2n} (Q_{,l_1}-Q_{,l_2})\in\mathbb{Z}.
\end{align}
In fact these are the necessary and sufficient conditions a  monopole wall parameters have to satisfy for such a monopole wall to exist.

Another  significant question is establishing criteria for when two spaces
${\cal M}_{Q_1,M_1,p_1,q_1,{\bf r}_1}$ and ${\cal M}_{Q_2,M_2,p_2,q_2,{\bf r}_2}$ are isometric.
For example, we shall describe an action of $SL(2,\mathbb{Z})$ on $(Q,M,p,q)$ that
identifies such isometric moduli spaces.  Moreover, this $SL(2,\mathbb{Z})$ group
acts on the monopole walls, as one might expect, mapping one solution to another up
to a gauge transformation.   In particular, the element
$S=\left(\begin{smallmatrix} 0 &-1\\1&0 \end{smallmatrix}\right)\in SL(2,\mathbb{Z})$
is the Nahm transform
\footnote{To be exact, it is a reflection of $y$ and $z$ coordinates followed by the Nahm transform that is the $S$ element of $SL(2,\mathbb{Z}).$}.
The action of a general element
$g=\left(\begin{smallmatrix} a & b\\c & d \end{smallmatrix}\right),\ ad-bc=1,$ is 
\begin{equation}
g: (Q,M,p,q)\mapsto \left(\frac{a Q+b}{c Q+d},\frac{M}{c Q+d}, 
      \frac{p}{c Q+d}, \frac{q}{c Q+d}\right).
\end{equation} 

We expect the moduli spaces of monopole walls to be of ALH type, {\sl i.e.}\ the asymptotic
volume growth of a ball of geodesic radius $R$ in a $4k$-dimensional moduli space is slower than $R^{2k}$.
We defer the study of their geometry, and focus here on the question of dimension of
the monopole-wall moduli spaces and the action of the modular group on them.  We find
that the most illuminating approach to all these questions is via the spectral
description of monopole walls of Section~\ref{Sec:Spec}. Before focussing on the problem
at hand, let us first discuss its relation to gauge and string theory.

\section{String Theory Dualities}\label{Dualities}
String theory was instrumental in identifying monopoles with vacua of quantum gauge theory \cite{Chalmers:1996xh}.  A configuration of $k$ SU(2)  BPS monopoles can be realized \cite{hep-th/9608163}  by suspending $k$ D-branes between a pair of parallel D-branes of two dimensions higher.  In our case, we consider $k$ D3-branes between a pair of parallel  D5-branes \cite{hep-th/9608163,Chalmers:1996xh}.  The effective theory on the pair of the D5-branes is the U(2) Yang-Mills, and the existence of the suspended D3-branes implies the presence of a monopole charge equal to $k$.  As the brane configuration respects eight real supercharges, the Yang-Mills configuration is BPS and satisfies the Bogomolny equation.  The equivalent description of the S-dual configuration \cite{Chalmers:1996xh, Hanany:1996ie} in terms of the theory on the D3-branes is via the supersymmetric quantum gauge theory in three dimensions, or its refinement --- four-dimensional theory with impurities \cite{Cherkis:2011ee}.  In this interpretation, it is a vacuum of such a theory that corresponds to a monopole solution.

Let us now identify a similar brane configuration that describes monopoles that are doubly-periodic.  Various string theory dualities allow us to relate it to quantum gauge theories and to the M theory five-brane on a spectral curve.  Table \ref{Table:Dualities} presents the schematic relation between the various string and M-theory configurations we describe below.


\begin{table}[h]
\begin{center}
\makebox[\textwidth]{
\tabcolsep=0.13cm
\begin{tabular}{ccc}
&
&
\hfill\shortstack{\footnotesize M5-brane wrapped\\ \footnotesize  on $\Sigma_M\subset\mathbb{C}^*\times\mathbb{C}^*$.}
\\
&
&
7.	\begin{tabular}{l|ccccccccccc} 
	  M    & 0 &1 &2 &3 &\cir{4}&\cir{5}&6 &7 &8 &9 &\raisebox{.6pt}{\textcircled{\raisebox{-.4pt} {$\mbox{\fontsize{9}{10}\selectfont $10$}$}}} \\ \hline
	{\footnotesize 2} M5 & x &x &x &x &x         & x       &   &   &    &   &    \\
	M5 & x &x &x &   &            &x       & x &  &    &   &x\\ 
	\end{tabular}

\\
1. 
\begin{tabular}{l|cccccccccc} 
  IIB    & 0 &1 &2 &3 &\cir{4}&\cir{5}&6 &7 &8 &9  \\ \hline
{\footnotesize 2} D5 & x &x &x &x &x         & x       &   &   &    &         \\
{\footnotesize k}  D3 & x &x &x &   &            &         & x &  &    &  \\ 
\end{tabular}
&
$\begin{array}{c}
T_5 S\\
\searrow
\end{array}$
&
\hspace{20pt}  $\downarrow S^1_M=S^1_{10}$ \hfill
\\
\hspace{20pt} $\downarrow T_5$ \hfill \shortstack{\footnotesize  Hitchin system with\\ \footnotesize ``group-valued Higgs field".}
&
&
4. \begin{tabular}{l|cccccccccc} 
 IIA     & 0 &1 &2 &3 &\cir{4}&\cir{5}&6 &7 &8 &9  \\ \hline
{\footnotesize 2} NS5 & x &x &x &x &x         & x       &   &   &    &         \\
{\footnotesize k} D4 & x &x &x &   &            &  x       & x &  &    &  \\ 
\end{tabular}
\\
2. \begin{tabular}{l|cccccccccc} 
 IIA     & 0 &1 &2 &3 &\cir{4}&\cir{5}&6 &7 &8 &9  \\ \hline
{\footnotesize 2} D4 & x &x &x &x &x         &        &   &   &    &         \\
{\footnotesize k} D4 & x &x &x &   &            & x         & x &  &    &  \\ 
\end{tabular}
& 
&
\hspace{20pt} $\downarrow T_4$ \hfill \shortstack{\footnotesize 5D $SU(k)$ Quantum  \\ \footnotesize Gauge Theory on $\mathbb{R}^3\times T^2$.}

\\
\hspace{20pt} $\downarrow T_4$ \hfill \shortstack{\footnotesize Nahm tranformed\\ \footnotesize monopole wall.}

&
&
5. \begin{tabular}{l|cccccccccc} 
  IIB    & 0 &1 &2 &3 &\cir{4}&\cir{5}&6 &7 &8 &9  \\ \hline
{\footnotesize 2} NS5 & x &x &x &x &x         & x       &   &   &    &         \\
{\footnotesize k} D5 & x &x &x &   & x           & x        & x &  &    &  \\ 
\end{tabular}
\\

3. \begin{tabular}{l|cccccccccc} 
  IIB    & 0 &1 &2 &3 &\cir{4}&\cir{5}&6 &7 &8 &9  \\ \hline
{\footnotesize 2} D3 & x &x &x &x &          &         &   &   &    &         \\
{\footnotesize k} D5 & x &x &x &   &x           &x       & x &  &    &  \\ 
\end{tabular}
&

&
\hspace{20pt} $\downarrow S$  \hfill \shortstack{\footnotesize 5D $SU(2)$ Quantum  \\ \footnotesize Gauge Theory on $\mathbb{R}^3\times T^2$.}
\\

&
$\begin{array}{c}
\nwarrow \\
S T_{45}
\end{array}$
&
6. \begin{tabular}{l|cccccccccc} 
  IIB    & 0 &1 &2 &3 &\cir{4}&\cir{5}&6 &7 &8 &9  \\ \hline
{\footnotesize 2} D5 & x &x &x &x &x         & x       &   &   &    &         \\
{\footnotesize k} NS5 & x &x &x &   & x      & x        & x &  &    &  \\ 
\end{tabular}
\end{tabular}
}
\caption{A circle of string theory dualities.}
\label{Table:Dualities}
\end{center}
\end{table}

\begin{enumerate}
\item For our purposes, we begin by considering Type IIB string theory in a space-time with two periodic directions, say the fourth and fifth.  We place a pair of D5-branes with $\mathbb{R}^{1,2}\times\mathbb{R}\times S_4^1\times S_5^1$ world-volumes separated along the sixth direction\footnote{$S^1_4$ and $S^1_5$ denote circles in directions $4$ and $5$ respectively.}; then we suspend $k$ parallel D3-branes on an interval $I_6$ in the sixth direction between them, so that D3-branes' world-volumes are $\mathbb{R}^{1,2}\times I_6.$  This brane configuration preserves eight supersymmetries, and thus its effective description in terms of the SU(2) gauge theory on the D5-branes is a BPS configuration, namely a BPS monopole on $\mathbb{R}\times S_4^1\times S_5^1.$ 

There are a number of string dualities that can be applied to this configuration, each producing a new interesting object dual to the $k$ doubly-periodic monopole configuration.  Let us explore some of these.

\item T-duality in the fifth direction produces a system of intersecting D4-branes: two of the D4-branes' world-volumes are $\mathbb{R}^{1,2}\times\mathbb{R}\times S_4^1$; and $k$ of the D4-branes, are  $\mathbb{R}^{1,2}\times S^1_5\times I_6.$  For a generic monopole wall these two sets of D4-branes fuse into a single four-brane with worldvolume $\mathbb{R}^{1,2}\times\Sigma_5$, with the curve $\Sigma_5\subset\mathbb{C}^*_{3,4}\times\mathbb{C}^*_{5,6}$ covering $\mathbb{C}^*_{3,4}=\mathbb{R}\times S^1_4$ twice and $\mathbb{C}^*_{5,6}=\mathbb{R}_6\times S^1_5$ $k$ times\footnote{The effects of brane bending that promote $I_6$ to $\mathbb{R}_6$ here and in the following are explained in detail in \cite{Witten:1997sc}.}.

Compared to this brane configuration, the periodic monopoles studied in \cite{Cherkis:2000ft} had the fourth direction noncompact, and that configuration was described (in terms of the theory on the $k$ parallel D4-branes) as a rank-$k$ Hitchin system.  Now, this brane configuration can roughly be viewed as a Hitchin system with ``a gauge-group-valued Higgs field'' or as a loop-group Hitchin system.

\item\label{NahmBranes} Next, applying subsequent T-duality in the fourth direction we have two D3-branes with world-volumes $\mathbb{R}^{1,2}\times \mathbb{R}$ and $k$ D5-branes stretching along $\mathbb{R}^{1,2}\times S_4^1\times S_6^1\times \mathbb{R}_6$.  From the point of view of the theory on the D5-branes, this is a configuration of two monopoles in U($k$) gauge theory.  We interpret it as the Nahm transform of the original $k$-monopole configuration.

\item\label{IIA} 
Now we follow a different path of dualities.  Starting back from the initial configuration 1, we apply S-duality followed by T-duality in the fifth direction.  The resulting Type IIA brane configuration consists of two NS5-branes with world-volumes $\mathbb{R}^{1,2}\times\mathbb{R}\times S^1_4\times S^1_5$ and $k$ D4-branes along $\mathbb{R}^{1,2}\times S^1_5\times \mathbb{R}_6.$  The main reason this configuration is useful is that it is directly related to an M-theory configuration (see item 7 below) with a single M5-brane on a curve\footnote{In fact $\Sigma_M=\Sigma_5,$ as explained in item 7.}  $\Sigma_M.$  This curve will play central role in the following discussion, and in our forthcoming computation of the asymptotic metric on the monopole moduli space.  We shall return to this configuration at the end of this section.

\item Applying T-duality along the fourth direction to our last configuration, we have two NS5-branes of the form $\mathbb{R}^{1,2}\times \mathbb{R}\times S^1_4\times S^1_5$ and $k$ D5-branes along $\mathbb{R}^{1,2}\times S^1_4\times S^1_5\times I_6.$  The effective low-energy theory on the D5-branes is the five-dimensional quantum gauge theory with the U($k$) gauge group and space-time $\mathbb{R}^{1,2}\times S^1_4\times S^1_5.$  In a way, in far infrared this is a three-dimensional super-Yang-Mills theory with ${\cal N}=4.$  (The remnant of the two compact directions is that the expectation values of two of the Higgs fields of the three-dimensional effective theory are the vacuum expectation values of the holonomies along the two periodic directions of the original five-dimensional theory.  These two holonomies and the dual photon constitute periodic directions in the moduli space.  Thus $3k$ out of $4k$ directions are expected to be periodic.  This is the reason to expect an ALH-type\footnote{See \cite{arXiv:1007.0044} for the definition of the ALE, ALF, ALG, and ALH nomenclature.} moduli space.)  This configuration identifies the initial monopole wall with a vacuum of a five-dimensional quantum gauge theory on $\mathbb{R}^{1,2}\times S^1\times S^1;$ while  the moduli space of this monopole wall is identified with the Coulomb branch of vacua of this five-dimensional quantum theory.  Such five-dimensional quantum theories on a two-torus were studied in \cite{arXiv:1107.2847}, in fact the gauge theory computation of \cite{arXiv:1107.2847} verifies that these spaces are indeed ALH. 

\item The S-dual of the last configuration is that of two D5-branes of the form $\mathbb{R}^{1,2}\times I_3\times S^1_4\times S^1_5$ and $k$ NS5-branes with world-volumes $\mathbb{R}^{1,2}\times S^1_4\times S^1_5\times \mathbb{R}_6.$  

Before we continue, we note that applying T-duality in the fourth and fifth directions followed by S-duality brings us back full circle to configuration \ref{NahmBranes}.

\item Let us now return to the Type IIA configuration \ref{IIA}.  Its M-theory lift is a set of $k$ M5-branes with world-volumes $\mathbb{R}^{1,2}\times S^1_5\times I_6\times S^1_{10}$ and two M5-branes along $\mathbb{R}^{1,2}\times \mathbb{R}\times S^1_4\times S^1_5.$  A special monopole configuration indeed corresponds to this five-brane intersection. A general monopole configuration, however, corresponds to a smooth curve that is a deformation of this intersection.  Namely, a curve $\Sigma_M\subset\mathbb{C}^*_{3,4}\times\mathbb{C}^*_{6,10}$ is a deformation of a reducible curve with two $\mathbb{C}^*_{3,4}$ components and $k$ $\mathbb{C}^*_{6,10}$ components.  The smooth M5-brane's world-volume is $\mathbb{R}^{1,2}\times S^1_5\times\Sigma_M.$

Since both 1 and 4 are related by T-duality $T_5$ to type IIB configurations that are S-dual, and since in M-theory, as in 7, compactified on a torus, S-duality amounts to interchanging the roles of the two circles of that torus, the type IIA configuration of 2 is the compactification of 7 with the fifth direction chosen as the M-theory direction $S^1_M=S^1_5.$  This implies that the curve $\Sigma_5$ on which  the D4-brane is wrapped is the same as the curve $\Sigma_M$ on which the M5-brane of 7 is wrapped.
\end{enumerate}

It is worth noting that a $T_4 S$ transformation of the original configuration 1, followed by a lift to M-theory, produces an M5-brane on $\mathbb{R}^{1,2}\times S^1_4\times\Sigma_4$ with a different curve $\Sigma_4$.  The two curves $\Sigma_4$ and $\Sigma_5$ will be exactly the two spectral curves $\Sigma_x$ and $\Sigma_y$ appearing in the spectral approach of Section \ref{Sec:Spec}.  The two M-theory configurations are related by $T_{4,5}$ duality, which is an interesting manifestation of M-theory T-duality taking M5-brane on $S^1_5\times \Sigma_5$ to that on $S^1_4\times\Sigma_4.$

The circle of string theory dualities we considered allows us to interpret $k$
doubly-periodic monopoles in SU(2) as 
\begin{itemize}
\item Vacua of five-dimensional supersymmetric quantum gauge theory with two periodic directions.  This theory can be viewed as a higher-dimensional version of the Seiberg-Witten theory. The infrared dynamics of this theory is given by a three-dimensional sigma-model with target space being the monopole-wall moduli space that we discussed in Section~\ref{Sec:Moduli}.
\item A single M-theory five-brane wrapped on a curve $\Sigma\subset\mathbb{C}^*\times\mathbb{C}^*.$  This curve plays an instrumental role in our spectral description of Section~\ref{Sec:Spec}.  
\item Two doubly-periodic U($k$) monopoles which are the result of the Nahm transform that we discuss in Section~\ref{Sec:Nahm}.
\end{itemize}

\section{Spectral Approach}\label{Sec:Spec}
As observed in \cite{Donaldson}, the three equations constituting the Bogomolny
equation~\eqref{Bogomolny} can be written as one complex and one real equation
\begin{equation}\label{ComplexForm}
\begin{cases}
\big[D_z-\ii D_y, D_x+\ii \Phi\big]=0, \\
\big[D_z-\ii D_y, (D_z-\ii D_y)^\dagger\big]+
     \big[D_x+\ii \Phi,(D_x+\ii \Phi)^\dagger\big]=0. 
\end{cases}
\end{equation}
In fact for any choice of direction $\hat{n}$ in the covering space of
$T^2\times\mathbb{R}$, we can write the Bogomolny equation as a pair consisting
of a complex equation and a real one, where the complex equation states that the
holomorphic covariant derivative in the plane orthogonal to $\hat{n}$ commutes
with the derivative $D_{\hat{n}}+\ii\Phi.$  One can use this equation to define
some spectral data, as we do below for $\hat{n}$ directed along the $x$- or $y$-axis.
In particular, there is an $SL(2,\mathbb{Z})$ worth of spectral descriptions,
each corresponding to a choice of $\hat{n}$ along any one of the generators of
the torus $T^2$. Below we formulate only two of these spectral descriptions.
We would like to emphasize that this $SL(2,\mathbb{Z})$ is the modular group of
the spatial torus $T^2$ acting on various spectral descriptions of the {\em same}
monopole wall, and that it is different from the $SL(2,\mathbb{Z})$ of
Section~\ref{Sec:Nahm} acting on monopole walls.

\subsection{Spectral data}
\subsubsection{$x$-spectral data}
Associated with any doubly-periodic solution, there is a set of
$x$-spectral data, as follows.
Let $V_x$ be defined by integrating $(D_x+\ii\Phi)\psi=0$ around one period
in the $x$-direction; so $V_x(y,z)$ takes values in the complexification
of the gauge group. In some cases, such as for the gauge group U(1), the fields
cannot be explicitly periodic in both $x$ and $y$ --- they are only periodic
up to a gauge transformation; in such cases, for computing $V_x$ one
should use a gauge in which all the fields are explicitly $x$-periodic.
Then the characteristic polynomial $F_x=\det[V_x(y,z)-t]$ of $V_x$ is
gauge-invariant, periodic in $y$, and holomorphic in $z-\ii y$. The
holomorphicity follows from the Bogomolny equation, namely from the first
equation in \eqref{ComplexForm}: $[D_z-\ii D_y , D_x+\ii\Phi]=0$,
by a straightforward adaptation of the argument in \cite{hep-th/0006050}.
So $F_x$ is a holomorphic (or meromorphic, if the field has singularities)
function of $s=\exp[2\pi(z-\ii y)]$, and it is a polynomial in $t$.
The vanishing of $F_x(s,t)$ defines a spectral curve
$\Sigma_x$, which lives in $\CC^*\times\CC^*$, where $s$ belongs to the first
$\CC^*$ factor and $t$ to the second. Since each point of $\Sigma_x$
corresponds to an eigenspace of $V_x$, we also get a coherent sheaf $M_x$
on $\Sigma_x$.
The stalks of $M_x$ are the corresponding eigenspaces, and are (at a general
point of $\Sigma_x$) one-dimensional; if $\Sigma_x$ is a Riemann surface,
{\sl i.e.}\ if it has no singularities, then $M_x$ is a holomorphic line bundle
\cite{hep-th/0006050}. The pair $(\Sigma_x, M_x)$ constitutes the
{\em $x$-spectral data} of the monopole field.

Given the boundary conditions of Sections~\ref{Sec:ABC} and \ref{Sec:Sing},
the function $F_x(s,t)$ is a degree $n$ polynomial in $t$, and its coefficients
are rational functions in $s$.  It is convenient to multiply $F_x(s,t)$ by
a common denominator $P(s),$ which is a polynomial in $s$, and to define the
{\em spectral polynomial} 
\begin{equation}
G_x(s,t)=P(s) F_x(s,t).
\end{equation}
This is a polynomial in both $s$ and $t$. We choose to normalize it so that
$(-1)^nP(s),$ which is the coefficient of $t^n,$ is a monic polynomial in $s$.

The pair $(\Sigma_x, M_x)$ is equivalent to the whole solution, and the
$4l$ real moduli of a monopole wall can be viewed as consisting of $2l$
moduli parametrizing the family of the spectral curves $\Sigma_x$ and
$2l$ moduli parametrizing line bundles $M_x$ over $\Sigma_x$.  In this
view, the moduli space is fibered by $2l$-dimensional tori over the moduli space of spectral curves.   

\subsubsection{$y$-spectral data}
Similarly, one obtains a matrix function $V_y$ by integrating
$(D_y+\ii\Phi)\psi=0$ around one period in the $y$-direction.  Here one has
to use a gauge in which the fields are periodic in the $y$-direction. 
The eigenvalues of $V_y$ form a spectral curve $\Sigma_y$ defined by
$\det[V_y(\st)-\tilde{t}]=0$ where $\st=\exp[2\pi(z+\ii x)]$, and $y$-spectral data.
The $x$-spectral data and the $y$-spectral data are related ---
in other words, not independent of each other --- as we shall see later.
The direct map between the $x$-spectral data $(\Sigma_x, M_x)$ and the
$y$-spectral data $(\Sigma_y, M_y)$, however, remains a mystery. 

\subsubsection{$z$-spectral data}
Finally, as in \cite{Donaldson}, there are $z$-spectral data associated
with the scattering problem
\begin{equation} \label{z-scatteringeqn}
    (D_z+\ii\Phi)\psi=0
\end{equation}
in the $z$-direction. One version of this, in the case of rank $n=2$,
is as follows. Choose a solution $\psi$ of (\ref{z-scatteringeqn}) which
satisfies $\psi\to0$ as $z\to\infty$, and which is
holomorphic in the sense that $(D_x+\ii D_y)\psi=0$. Note that
$[D_z+\ii\Phi, D_x+\ii D_y]=0$ from the Bogomolny equations, so this
holomorphicity requirement is consistent. Next choose a holomorphic
solution $\psi_{-}$ which does not vanish as $z\to-\infty$.
Define a function $B(x,y)$ by
\[
  \psi(x,y,z)=B(x,y)\, \psi_{-}(x,y,z) +
         \mbox{part which vanishes as $z\to-\infty$}.
\]
Then $B$ is holomorphic in $\zeta=x+\ii y$. Its zeros are the
{\em spectral points},
labelling the $z$-lines {\em (spectral lines)}\ along which there is a
solution $\psi$ of (\ref{z-scatteringeqn}) with $\psi\to0$ as $z\to\pm\infty$.
Of course, $B$ depends on the choice of $\psi$ and $\psi_{-}$, but
the fact that the field has a standard asymptotic form enables one to make
a natural choice. In any event, making a different choice has the effect
of multiplying $B$ by a nowhere-zero holomorphic function, which does not
affect its zeros and hence the spectral points.
Similarly, in the general U($n$) case, one may define $z$-spectral data,
along the lines of \cite{Jarvis1, Jarvis2}.

\subsubsection{Examples}
Let us now turn to some examples, and compute their spectral curves.  Anticipating
our discussion of the Nahm transform in Section~\ref{Sec:NTransf}, we mention
that it corresponds to interchanging the $s$ and $t$ variables.  Thus we pay
particular attention to the action of this symmetry on the spectral curves we discuss.

For the $Q_{\pm}=1, M=p=0$ case of the constant-energy solution in
Eq.~\eqref{ConstE} above, we clearly have $V_x(s)=s$; and the
corresponding spectral curve $s=t$ is invariant under the interchange of $s$
and $t$ --- the underlying reason for this is that the constant-energy solution
maps to itself under the Nahm transform \cite{Ward:2005nn}. Slightly more generally,
for the constant-energy solution
Eq.~\eqref{ConstE} with $Q_{\pm}=Q>1, M=p=0$,
we get $V_x(s)=s^Q$. The Nahm transform $(\Phi',A'_j)$ can again be computed
explicitly in terms of theta-functions, as in the $Q=1$ case, and
it turns out to be a diagonal U($Q$) field; in particular,
$\Phi'=(2\pi\ii z/Q){\bf I}_Q$,
where ${\bf I}_Q$ is the identity $Q\times Q$ matrix.

For the Dirac 1-pole wall of
Eqs.~(\ref{dirac_phi1},\ref{dirac_gaugepotplus},\ref{dirac_gaugepotminus}),
we get $V_x(s)=a(s-1)$, where $a=\ee^{2\pi M_+}$ is
determined by $\phi_0$. The corresponding spectral curve is mapped
under $s\leftrightarrow t$ to $V_{x'}(s)=t$ with $V_{x'}(s)=a^{-1}s +1$:
this corresponds to the fact that
the Nahm transform of a Dirac 1-pole wall is another Dirac 1-pole wall.
If the pole is located at $\vr_+=(r_1,r_2,r_3)$ rather than at the origin,
then we get $V_x(s)=a(s-\beta)$ and $V_y(\st)=a(s-\tilde{\beta})$,
where $\beta=\exp[2\pi(r_3-\ii r_2)]$ and
$\tilde{\beta}=\exp[2\pi(r_3+\ii r_1)]$. Note that $V_x(s)$ and $V_y(\st)$
are closely related, rather than being independent data (as was emphasized
previously).

Finally, for the 2-pole example of Eq.~\eqref{dirac_phi2}, $V_x$ has the form
$V_x(s)=A(s-B)(s-C)/s$.
In this case, the interchange $s\leftrightarrow t$ gives the spectral
curve $\widetilde{\Sigma}_{x'}$ with equation of the form
$\det[V_{x'}(s)-t]=0$, where $V_{x'}$ is a $2\times2$ matrix with
$\tr(V_{x'})=B+C+A^{-1}s$ and $\det(V_{x'})=BC$. This corresponds to
a U(2) system, with constant trace part.
If the original Dirac 2-pole system is centred, in other words
$\vr_{-1}=-\vr_{-2}$, then $BC=1$ and the Nahm-transformed system has
gauge group SU(2). This particular SU(2) solution will be described later.

\subsection{Newton Polygon}
As we outlined above, a monopole wall has spectral polynomials $G_x(s,t)$
and $G_y(s,t)$.  Each has its corresponding Newton polygon, denoted
respectively by $N_x$ and $N_y$.  We demonstrate shortly that in fact $N_x=N_y$.  

Considering a spectral curve, say $\Sigma_x\in{\mathbb C}^*\times{\mathbb C}^*$,
given by a polynomial relation $G_x(s,t)=0$, we would like to understand its
asymptotic behaviour as either $s$ or $t$ approaches $0$ or $\infty$. This
behaviour translates into conditions on the corresponding monopole fields.
There are three possibilities:
\begin{enumerate}
\item $t\rightarrow\infty$ while $s\rightarrow s_0$, or $t\rightarrow 0$ while
$s\rightarrow s_0$, with $s_0$  finite.  In these cases, the monopole has a
Dirac singularity, of respectively positive or negative type, positioned at
the point with $y$- and $z$-coordinates given by
$z-\ii y=\frac{1}{2\pi}\log(\bar{s}_0)$.
\item $s\rightarrow\infty$ or $s\rightarrow 0$ while $t\rightarrow t_0$, with
$t_0$ finite. In this case, the real and imaginary parts of $\log(t_0)$ are
the constant asymptotic values of eigenvalues of the Higgs field, and of the
holonomy around the $x$-direction, as $z\rightarrow+\infty$ or $-\infty$
respectively. 
\item For some relatively prime integers $\alpha$ and $\beta$, with $\beta$
positive, we have $t\sim s^{\alpha/\beta}$ as
$s\rightarrow\infty$ or as $s\rightarrow 0$. In this case, there are $\beta$
(or, more generally, a multiple of $\beta$) 
eigenvalues of the Higgs field with the dominant asymptotic behaviour
$2\pi \ii\frac{\alpha}{\beta} z$.
\end{enumerate}

This behaviour of the spectral curve can be conveniently read off from its
Newton polytope \cite{Newton}, which in our case is a Newton polygon.
A {\em Newton polygon}  $N$ of a polynomial $G(s,t)$ is constructed as follows.
For any monomial $s^a t^b$ which is present in $G(s,t)$ with nonzero coefficient,
we mark the point $(a,b)$ on an integer lattice. The Newton polygon $N$  is
a minimal convex polygon with lattice vertices containing all of the marked
points in it.  If $G(s,t)$ has degree $n$ in $t$ and degree $m$ in $s$,
then its Newton polygon fits into an $m\times n$ rectangle.

Given an edge $e$ of the Newton polygon, we denote by $G_e(s,t)$
a polynomial consisting of the terms in $G(s,t)$ that correspond to the
points belonging to the edge~$e$. Some of the edges of this polynomial can
lie on the edges of the ambient $m\times n$ rectangle. If such an edge
$e_N$ containing $r_{+0}+1$ points belongs to the top (northern) edge of the ambient
rectangle, then we have $r_{+0}$ branches  satisfying the condition~1 above.
The corresponding edge polynomial has the form $G_{e_N}(s,t)=P_N(s)t^n$ for
some polynomial $P_N(s)$. The $r$ nonzero roots $m_{+0}^\nu$ (with
$\nu=1,\ldots,r_{+0}$) of $P_N(s)$ give the positions
of the positive Dirac singularities. If they are all distinct, then these
are the basic Dirac singularities, with one of the eigenvalues of the
Higgs field unbounded above near the singularity.
Analogously, the edge $e_S$ containing $r_{-0}+1$ points belonging to the bottom
(southern) edge gives the positions $m_{-0}^\nu$ (with $\nu=1,\ldots,p_{-0}$)
and number $r_{-0}$ of points with negative Dirac singularities. 

An eastern edge $e_E$ of the Newton polygon has $G_{e_E}(s,t)=P_E(t)s^m$,
and corresponds to possibility~2 above, with the real and imaginary parts of the
logarithm of the nonzero roots of $P_E(t)$ being the limiting values of respectively
the eigenvalues of the Higgs field and logarithm of the holonomy
eigenvalues as $z\rightarrow+\infty$.
Analogously, the western edge $e_W,$ if it exists, gives the finite-limit eigenvalues
of the Higgs field and holonomy at $z\rightarrow-\infty$.

Since the spectral polynomial $G(s,t)$ has degree $(m,n)$, and is not
divisible by $s$ or by $t$, its Newton polygon always has some points on
each of the edges of the ambient rectangle.  Removing such points, and any
edges belonging to the edges of the ambient rectangle, leaves at most four
connected components.  We shall call these, according to their position,
North-West, North-East, South-East, and South-West components.  All edges
belonging to these components determine the third type of asymptotic Higgs
eigenvalue behaviour as $z\rightarrow\pm\infty$. In particular, the NE
component determines the eigenvalues of the Higgs field that grow linearly
with $z$ as $z\rightarrow+\infty$, while the SE components determine the
eigenvalues of the Higgs field that decay linearly with $z$ as
$z\rightarrow+\infty$. The SW and NW components determine, respectively,
the linearly decaying and growing components of the Higgs field as
$z\rightarrow-\infty$. 

Strictly speaking, these statements based on the Newton technique are all
about the asymptotic behaviour of the spectral curve $\Sigma_x$, and not about
the eigenvalues of the Higgs field.  In order to translate them into statements
about the Higgs field, we need to appeal to the geometry of the {\em amoebas}
corresponding to $F(s,t)=0$, and theorems of \cite{GKZ}.

\subsubsection{Amoebas}
The {\em amoeba} $A_x$ of the spectral curve $\Sigma_x$ is the image of
$\Sigma_x$ under the map 
\begin{align}
{\mathbb C}^*\times{\mathbb C}^*&\rightarrow{\mathbb R}^2\\
(s,t)&\mapsto(\log|s|, \log|t|).
\end{align}
An amoeba is a connected domain, with its complement consisting of connected
convex components.  Each such complement component can be associated to an integer point
of the Newton polygon (or its interior).  Nearby noncompact components are separated by the
amoeba's tentacles, which are exponentially narrow spikes heading to infinity.
Each tentacle asymptotes to a straight line orthogonal to an edge of the Newton polygon.
We illustrate this by looking
at some examples of monopole walls, their Newton polygons and amoebas.

\subsubsection{Examples}
Here we give some examples of amoebas and Newton polygons associated with the
monopole wall examples above.  The number of internal points of the Newton polygon
is the number of complex parameters which can be varied while keeping the
asymptotics of the spectral curve fixed.  Since the spectral pair $(\Sigma_x, M_x)$
determines the monopole wall, each internal point gives two real moduli of the
monopole-wall moduli space.  

\begin{figure}[htbp]
\begin{center}
\includegraphics[width=0.4\textwidth]{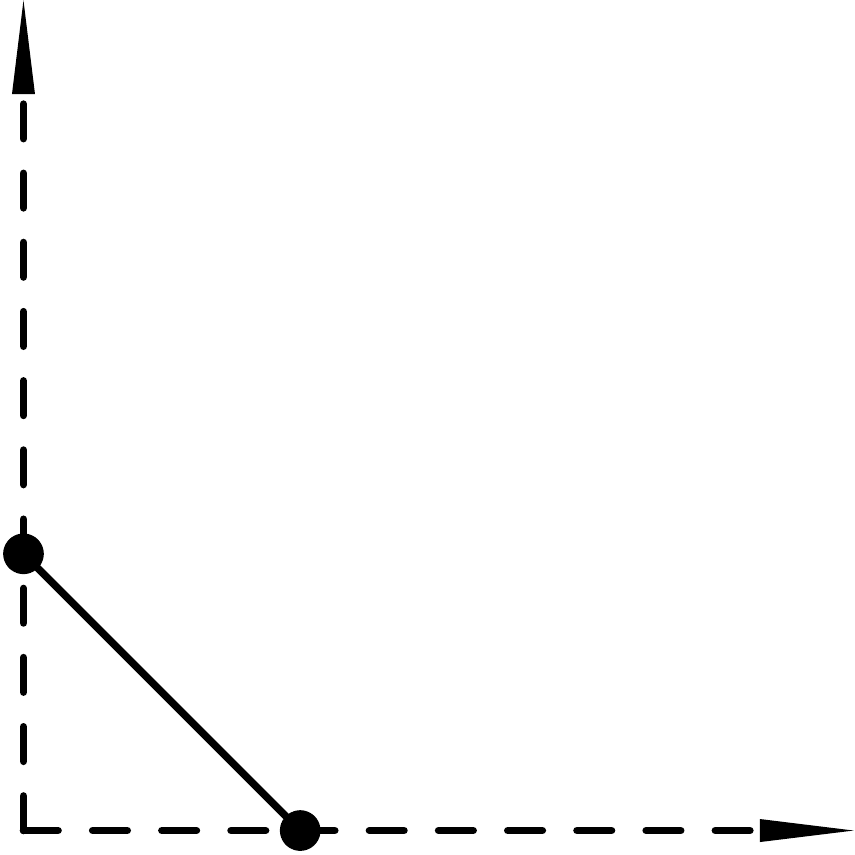}
\caption{Newton polygon for the constant-energy $U(1)$ solution.}
\label{Fig:ConstE}
\end{center}
\end{figure}

\bigskip\noindent{\bf Constant-energy field.}
In this case, the spectral curve is $\Sigma_x=\{(s,t)\, |\, s=t\},$ its
Newton polygon is in Figure~\ref{Fig:ConstE}, and a degenerate amoeba is given
by the main diagonal.
The Newton polygon is degenerate to an interval with endpoints $(0,1)$ and $(1,0)$ and has no internal points;
accordingly the spectral curve has no independent parameters, and is completely
determined by the data at infinity.

\begin{figure}[htbp]
\begin{center}
\includegraphics[width=0.4\textwidth]{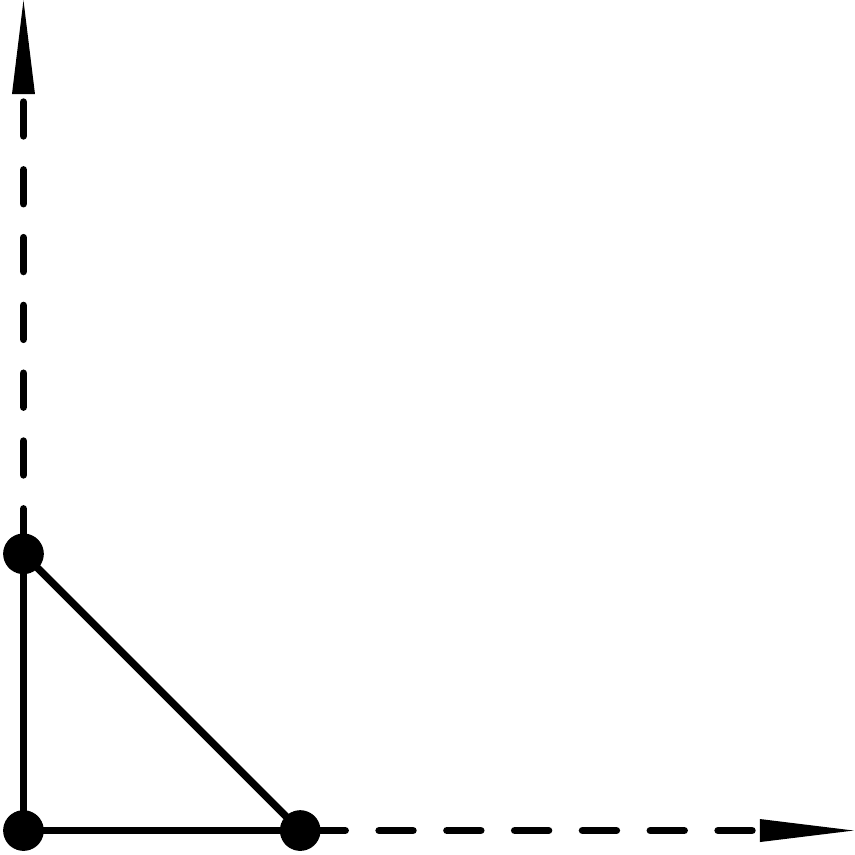}
\includegraphics[width=0.4\textwidth]{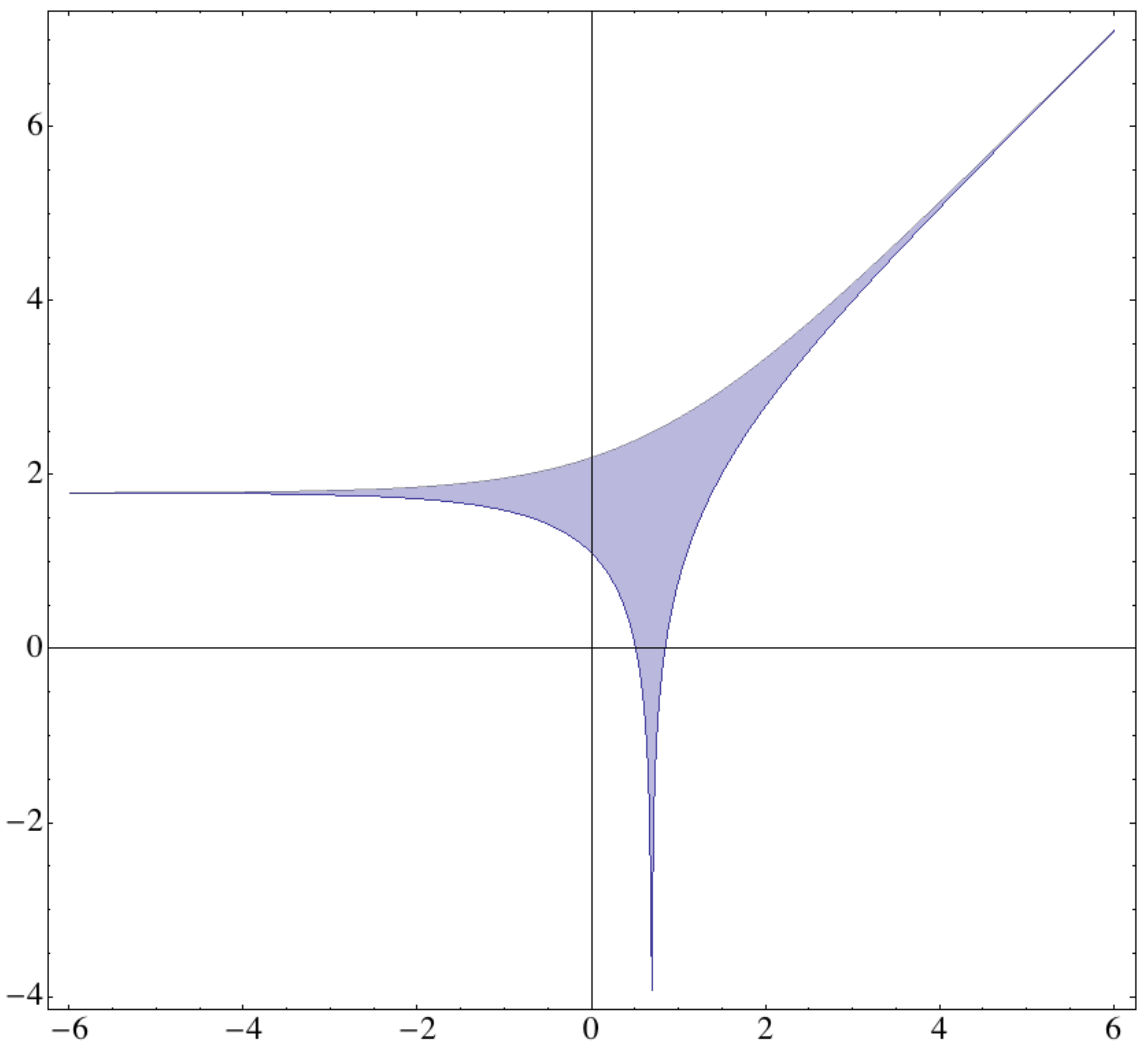}
\caption{Basic one-pole wall with the spectral curve $t-3s+6=0$: the vertical
tentacle indicates the position of a
negative pole, the left tentacle corresponds to $Q_-=0,$ and the NE orientation
of the right tentacle to $Q_+=1.$}
\label{Fig:1Wall}
\end{center}
\end{figure}

\bigskip\noindent{\bf Basic U(1) one-pole wall.}
In this case, the spectral curve is $\Sigma_x=\{(s,t)\, |\, a(s-1)=t\}$,
and the Newton polygon with an example of an amoeba are in Figure~\ref{Fig:1Wall}.
In the left-hand plot, the abscissa is the power of the $s$ variable and the
ordinate is that of the $t$ variable.  In the right-hand plot of the amoeba,
the axes are $\log|s|$ and $\log|t|$, and the shaded area corresponds to all
values of $(s,t)\in\Sigma_x.$  
For this solution, the spectral curve is again completely determined by
the boundary conditions.

\begin{figure}[htbp]
\begin{center}
\includegraphics[width=0.4\textwidth]{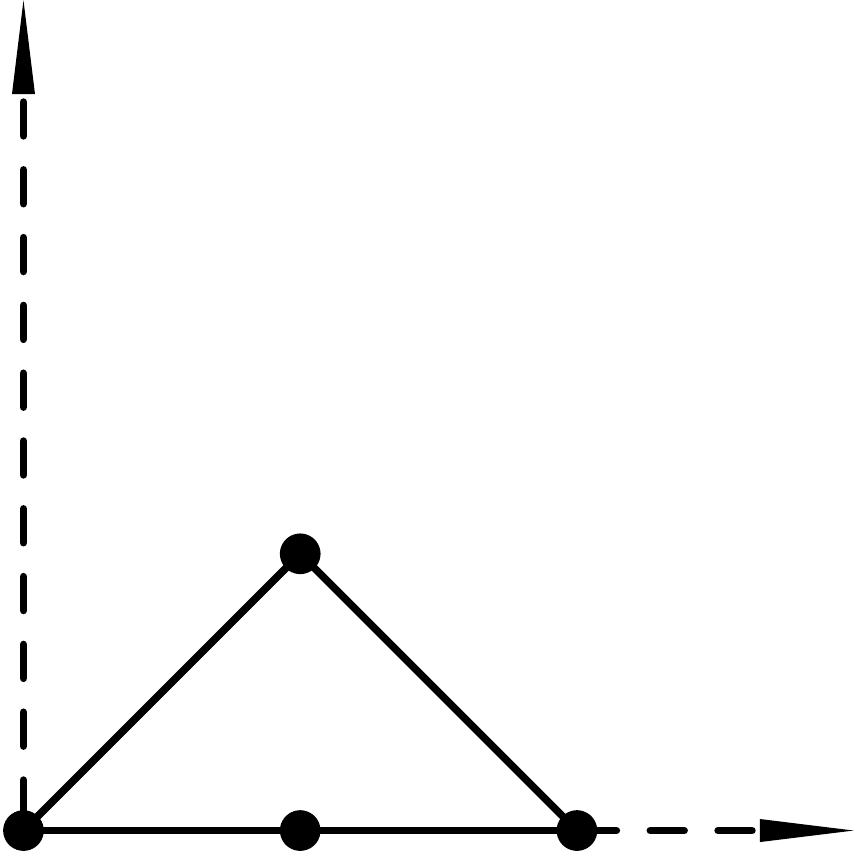}
\includegraphics[width=0.4\textwidth]{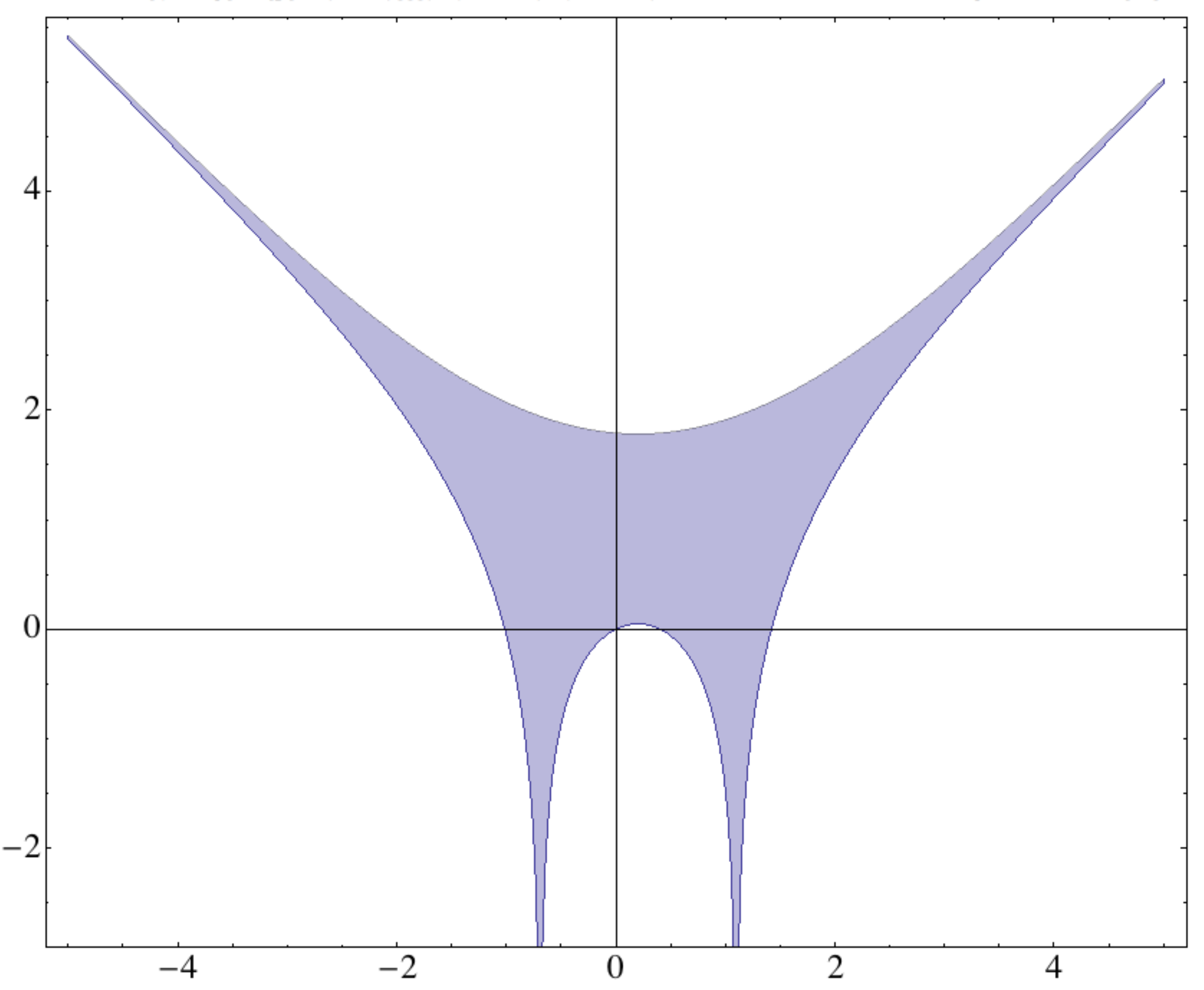}
\caption{A two-pole wall with the spectral surve $2ts=2s^2-7s+3$: the two vertical
tentacles indicates the two negative
poles, the left NW tentacle corresponds to $Q_-=1,$ and the NE orientation of
the right tentacle to $Q_+=1.$}
\label{Fig:2Wall}
\end{center}
\end{figure}

\bigskip\noindent{\bf Two-pole U(1) wall.}
The spectral curve is $\Sigma_x=\{(s,t)\, |\, (s-B)(s-C)=st/A \}$, and it corresponds
to the Newton polygon in Figure~\ref{Fig:2Wall}.  As in the two previous examples,
the Newton polygon contains no internal points, so this spectral curve has no
free parameters.

\bigskip\noindent{\bf SU(2) monopole wall with $(Q_-,Q_+)=(0,1)$.}
The spectral curve has the form $\Sigma_x=\{(s,t)\, |\, (t-a)(t-a^{-1})= st\}$.
Its Newton polygon and amoeba are given by figures similar to those of Figure~\ref{Fig:2Wall},
with the interchange of the abscissa and the ordinate axes. 

\begin{figure}[htbp]
\begin{center}
\includegraphics[width=0.4\textwidth]{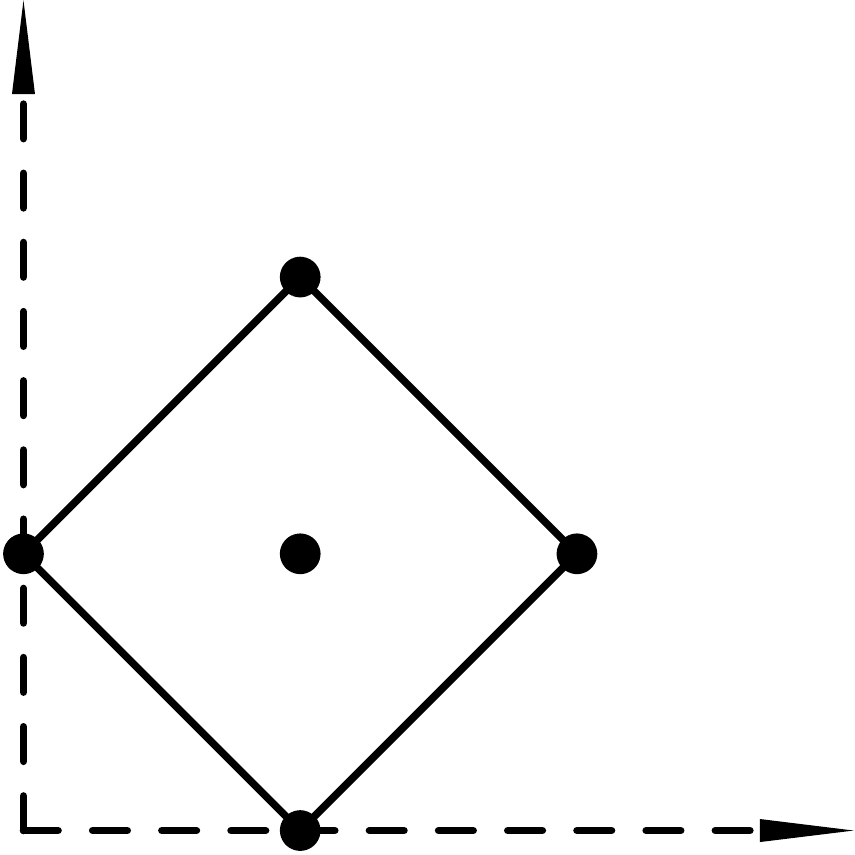}
\includegraphics[width=0.4\textwidth]{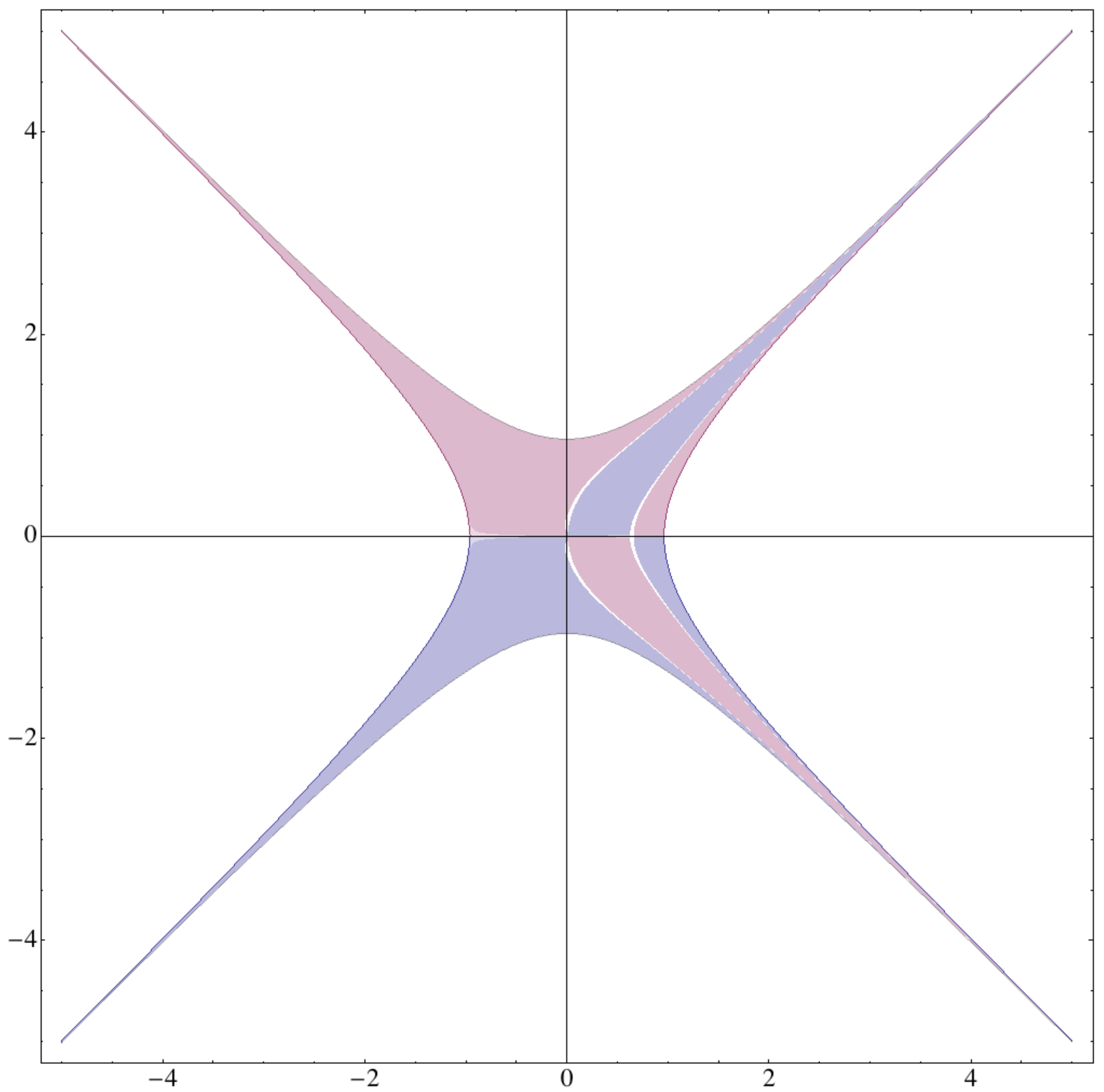}
\caption{The constant energy $SU(2)$ monopole wall has four-dimensional
moduli space. Two of the four moduli are given by the coefficient $a$ of the
monomial $st$ corresponding to the internal point in the Newton polygon.
This amoeba is for $a=1.$}
\label{Fig:SU2_Rombus}
\end{center}
\end{figure}

\begin{figure}[htbp]
\begin{center}
\includegraphics[width=0.4\textwidth]{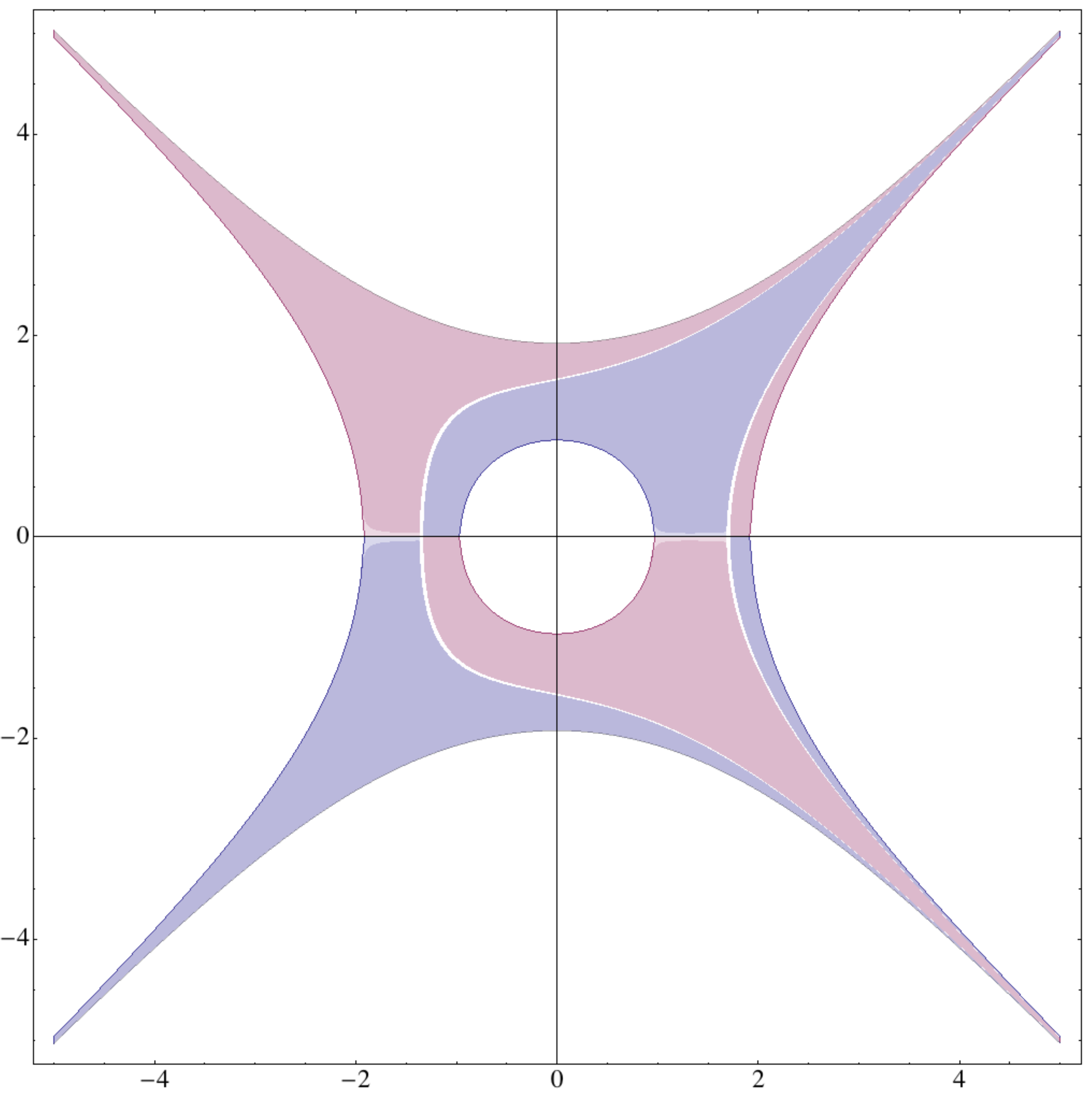}
\caption{The constant energy $SU(2)$ monopole amoeba for $a=5.$}
\label{Fig:SU2_R_Hole}
\end{center}
\end{figure}

\begin{figure}[htbp]
\begin{center}
\includegraphics[width=0.4\textwidth]{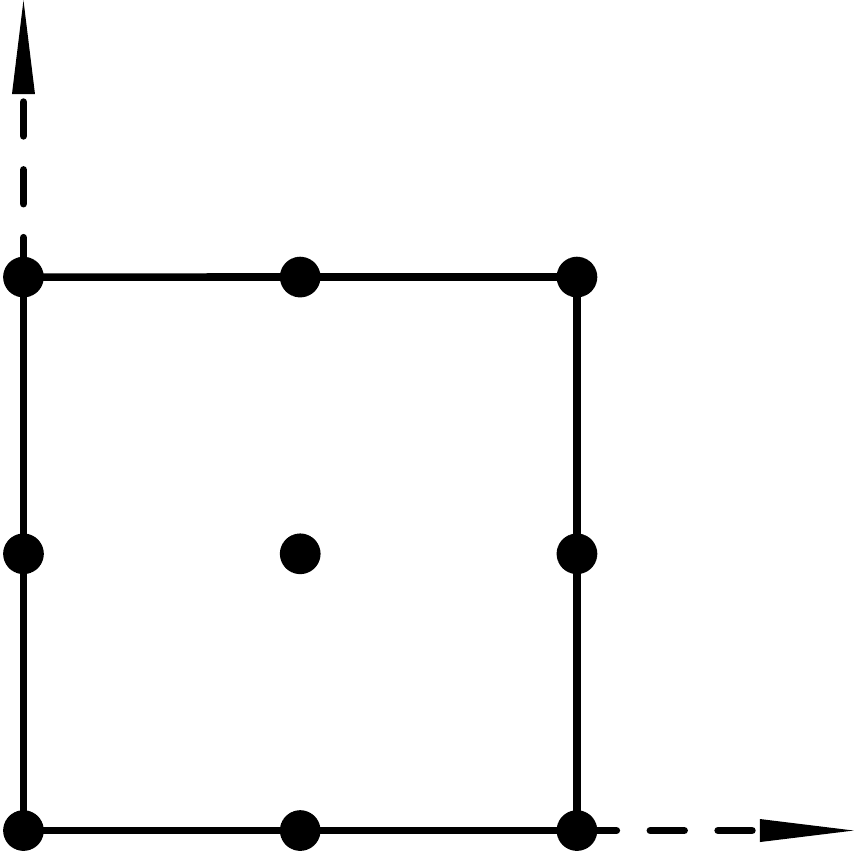}
\includegraphics[width=0.4\textwidth]{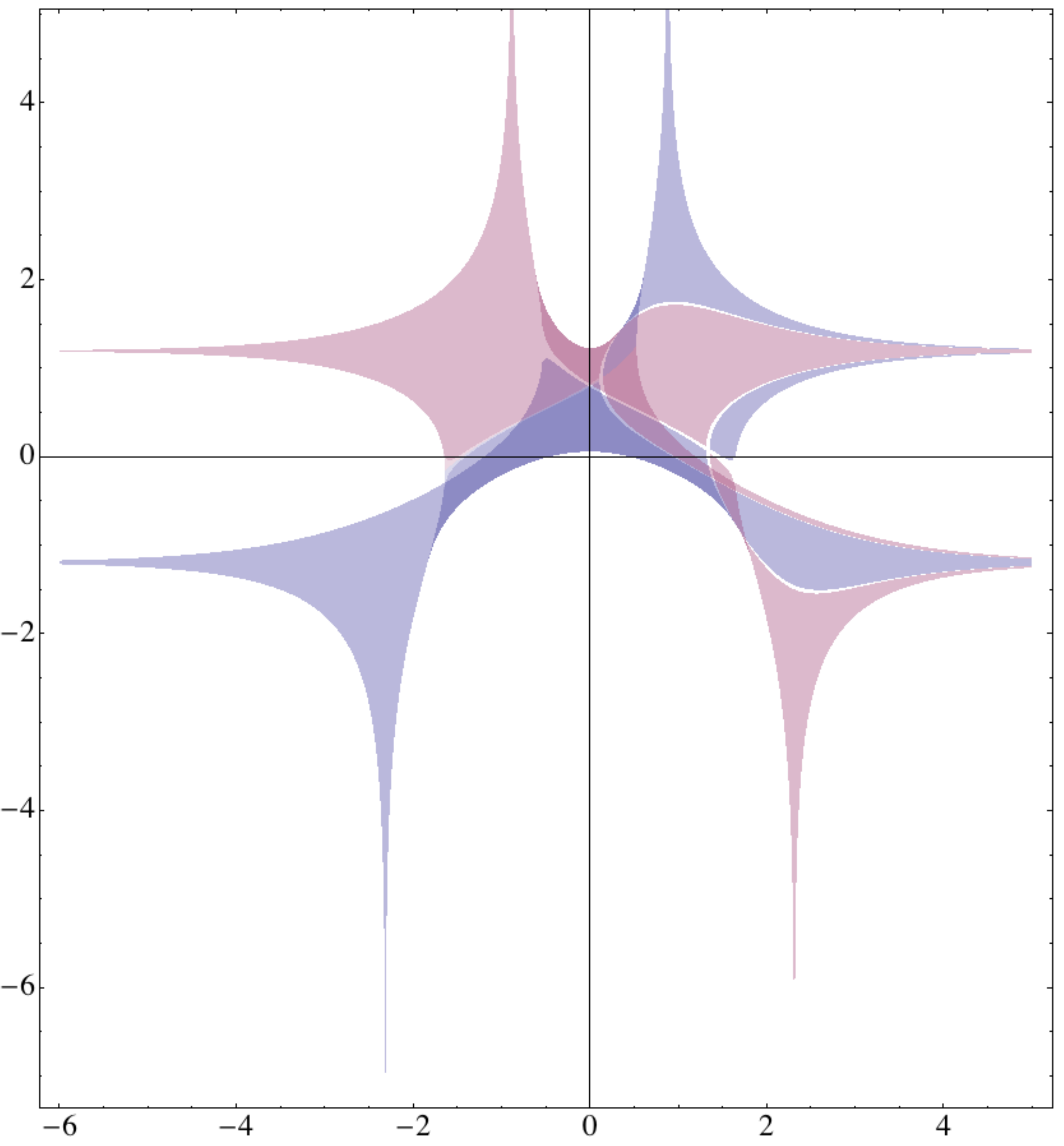}
\caption{A balanced $U(2)$ monopole wall with $r_{+0}=r_{-0}=2.$ The spectral
 curve in this example  is $(s^2-2s-1)t^2+3(s^2+s-1)t-s^2-10s+1=0.$}
\label{Fig:Balanced2}
\end{center}
\end{figure}

\bigskip\noindent{\bf SU(2) monopole wall with $(Q_-,Q_+)=(1,1)$.}
The spectral curve is $\Sigma_x=\{(s,t)\, |\, s t^2-s^2 t-t+s+a=0\}$.
Its Newton polygon and amoeba (for $a=1$) are given in Figure~\ref{Fig:SU2_Rombus}.
Various white lines and shadings appearing in this figure are artifacts of the graphing process
and should be ignored; the same applies to Figures~\ref{Fig:SU2_R_Hole}
and \ref{Fig:Balanced2}.
This is the first example where the Newton polygon contains an internal point.
Since there is one such point, the moduli space has at least two real dimensions.
In fact, as we argue below, it is four-dimensional. 
Note that the simplest choice of $a=0$, used later in
Eq.~\eqref{const-energy-W}, corresponds to a
degenerate solution with spectral curve $\{s=t\}\cup\{s=1/t\}$. Such a solution
is given by a superposition of two constant-energy U(1) solutions of the first
example. Changing the value of the coefficient corresponding to the internal point
changes the geometry of the amoeba.  As the value increases, the topology of the amoeba
can change, as Figure~\ref{Fig:SU2_R_Hole} illustrates; this is a general phenomenon.  

\bigskip\noindent{\bf A balanced U($n$) monopole.}
A monopole wall with all monopole charges vanishing is a particularly interesting
case we call a {\em balanced monopole}.  It has to have an equal number of positive
and negative singularities: $r_{-0}=r_{+0}.$  The positions of its negative and
positive singularities $\sigma_{-0}^{\nu}$ and $\sigma_{+0}^{\nu}$ are related to the
asymptotic values of the Higgs field and holonomy\footnote{See Eq.~\eqref{Eq:asigma} of the next example for the exact expressions for $a_{\pm1}$ and $\sigma_{\pm0}^\nu.$} $a_{-1}^\mu$ and $a_{+1}^\mu$  by
\begin{equation}
\prod_{\nu=1}^n\frac{ \sigma_{+0}^{\nu}}{\sigma_{-0}^{\nu}}=
   \prod_{\mu=1}^m \frac{a_{+1}^\mu}{a_{-1}^\mu}.
\end{equation}
A balanced U($n$) monopole wall with $r_{+0}=r_{-0}=m$ singularities has its
Newton polygons given by $m\times n$ rectangles.  An example of a Newton
polygon and amoeba of a U(2) monopole wall with $r_{+0}=r_{-0}=2$ is given in
Figure~\ref{Fig:Balanced2}.

\bigskip\noindent{\bf U(2) monopoles with four singularities and $Q_+=Q_-=(0,0)$.}
The simplest case of a balanced monopole with moduli has gauge group U(2). Consider
a monopole wall solution with two negatively charged Dirac
singularities  at $\vr_{-,1}=\vr_1$ and $\vr_{-,2}=\vr_3$ and two positively
charged Dirac singularities at $\vr_{+,1}=\vr_2$ and $\vr_{+,2}=\vr_4:$
\begin{equation}
\Phi=\left(\begin{array}{cc}
\frac{(-1)^\alpha}{2 |\vr-\vr_\alpha|} & 0 \\
0 & 0
\end{array}\right)+O(|\vr-\vr_\alpha|).
\end{equation}
The Higgs field is regular at infinity; let
\begin{eqnarray}
2\pi\ii (M_1, M_3)&=&\lim_{x_3\rightarrow-\infty}{\rm EigVal}\ \Phi,\\
2\pi\ii (M_2, M_4)&=&\lim_{x_3\rightarrow+\infty}{\rm EigVal}\ \Phi.
\end{eqnarray}
We denote the logarithm of the eigenvalues of the gauge field
monodromy $W(x_1,2\pi,x_3)$ around the $x$-direction at infinity by
$2\pi\ii p_1$, $2\pi\ii p_2$, $2\pi\ii p_3$ and $2\pi\ii p_4$:
\begin{eqnarray}
\lim_{x_3\rightarrow-\infty}{\rm EigVal}\ W(x_1,2\pi,x_3)
     &=&(e^{2\pi\ii p_1}, e^{2\pi\ii p_3}),\\
\lim_{x_3\rightarrow+\infty}{\rm EigVal}\ W(x_1,2\pi,x_3)
     &=&(e^{2\pi\ii p_2}, e^{2\pi\ii p_4}).
\end{eqnarray}
Let us combine these data into 
\begin{equation}\label{Eq:asigma}
a_j=\exp[2\pi (M_j+\ii p_j)]\quad {\rm and}\quad 
\sigma_j=\exp[2\pi\ii (r_j^1+\ii r_j^3)].
\end{equation} In terms of these, the
spectral curve has the behaviour
\begin{equation}\label{sing}
t\sim (s-\sigma_j)^{(-1)^j}\ {\rm as}\ s\rightarrow \sigma_j,
\end{equation}
and
\begin{equation}\label{ends}
t\rightarrow a_j\ {\rm as}\  s\rightarrow 0\ {\rm or}\ \infty.
\end{equation}

Since the spectral curve $\{F_x(s,t)=0\}$ is a double cover of the $s$-plane,
the function $F_x(s,t)$ is quadratic in $t$; and since we have two positive and two
negative Dirac singularities, $G_x(s,t)$ is quadratic in $s$.  The asymptotic
conditions (\ref{ends}) imply that $G_x(s,t)$ is proportional to
\begin{equation}
(t-a_1)(t-a_3) s^2+f(t) s+D(t-a_2)(t-a_4),
\end{equation}
with $D$ some constant and  $f(t)$ a quadratic polynomial in $t$.

The singularity structure (\ref{sing}) constrains $G_x(s,t)$ to be proportional to
\begin{equation}
(t-a_1)(t-a_3)s^2-\left((\sigma_2+\sigma_4)t^2-u t +
 a_1 a_3(\sigma_1+\sigma_3)\right)s+\frac{a_1 a_3}{a_2 a_4}\sigma_1\sigma_3(t-a_2)(t-a_4).
\end{equation}
Moreover, it implies the following relation between the asymptotics and the singularities:
\begin{equation}
a_1 a_3\sigma_1\sigma_3=a_2 a_4\sigma_2 \sigma_4.
\end{equation}
Here $u$ is the coordinate on the moduli space of the solutions.  The
whole moduli space can be thought of as an elliptic fibration over
the $u$-plane with fiber consisting of the Jacobian of $\Sigma_x$.
This description provides a natural complex structure $I$ on the
moduli space.


\section{Moduli and Asymptotics}

\subsection{Newton Polygon from the Boundary Data}
Consider a U($n$) monopole wall with $r_{-0}$ negative Dirac singularities and $r_{+0}$
positive Dirac singularities, and the spectrum of distinct charges
\begin{align}\label{Eq:Qs1}
Q_-&=\{Q_{-1},Q_{-2},\ldots,Q_{-f_-}\},& Q_{-1}&>Q_{-2}>\ldots>Q_{-f_-},\\
\label{Eq:Qs2}
Q_+&=\{Q_{+1},Q_{+2},\ldots,Q_{+f_+}\},& Q_{+1}&>Q_{+2}>\ldots>Q_{+f_+}.
\end{align}
The charges are rational, so we write $Q_{\pm j}=\alpha_{\pm j}/\beta_{\pm j}$
either with  $(\alpha_{\pm j}, \beta_{\pm j})=(0,1)$, or with
$\alpha_{\pm j}\in\mathbb{Z}$ and $\beta_{\pm j}\in\mathbb{N}$ relatively prime. 
The multiplicities of the respective eigenvalues are proportional to the denominators,
so they can be written as  $r_{-1}\beta_{-1},\ldots,r_{-f_-}\beta_{-f_-}$ and
$r_{+1}\beta_{+1},\ldots,r_{+f_+}\beta_{+f_+}$ for some positive integers
$r_{\pm j}.$  By construction, the rank of the bundle is
$n=\sum_{j=1}^{f_-} r_{- j}\beta_{- j}=\sum_{j=1}^{f_+} r_{+ j}\beta_{+ j}$.
Let us form {\em elementary} vectors
\begin{align}\label{ElementaryVec}
e_{-0}&=\left(\begin{matrix} -1\\ 0\end{matrix}\right),&
e_{-j}&=\left(\begin{matrix} -\alpha_{-j} \\ \beta_{-j}\end{matrix}\right),&
e_{+0}&=\left(\begin{matrix} 1\\ 0 \end{matrix}\right),&
e_{+j}&=\left(\begin{matrix} \alpha_{+j} \\ -\beta_{+j}\end{matrix}\right).
\end{align}
Then the edges in the sequence 
$$r_{-0}e_{-0}, r_{-1}e_{-1},\ldots, r_{-f_-}e_{-f_-},
   r_{+0}e_{+0}, r_{+1}e_{+1},\ldots, r_{+f_+}e_{+f_+}$$ 
form consecutive edges of the Newton polygon of this doubly-periodic monopole,
and an edge $r_j e_j$ contains $r_j+1$ integer points.

This picture makes it clear that the charges are such that the polygon closure conditions 
\begin{align}
\sum r_{-j}\beta_{-j}&=\sum r_{+j}\beta_{+j},&
r_{-0}+\sum r_{-j}\alpha_{-j}&=r_{+0}+\sum r_{+j}\alpha_{+j},
\end{align}
are satisfied as in Eq.~\eqref{Closedness}.

\subsection{Number of Moduli}
The ambient space ${\mathbb R}^2$ of the amoeba is dual
 to the plane in which the Newton polygon is defined.  A number of
useful theorems about amoebas can be found in \cite{GKZ}.
In particular, the perimeter integer points divide the edges into subedges,
and an amoeba $A_x$ generically has as many tentacles as there are subedges of the
Newton polygon of $G(s,t)$.  Each tentacle asymptotes to a line orthogonal
to the corresponding edge of the Newton polygon.
The number of holes in an amoeba, {\sl i.e.}\  the number of compact components
of its complement, is bounded above by the number of integer points inside
the Newton polygon.  As proved in \cite{MR}, the area of an amoeba is bounded
above by the area of the corresponding Newton polygon:
\begin{equation}\label{AmoebaIneq}
{\rm Area}(A)\leq\pi^2\,{\rm Area}(N).
\end{equation}
This inequality is saturated by Harnack curves.  A natural question to ask is whether
there is anything special about monopole walls which have both spectral curves
$\Sigma_x$ and $\Sigma_y$ being Harnack curves.  One interesting property of a
Harnack curve is that the number of holes in its amoeba equals the number of internal
points of its Newton polygon.  Moreover, a Harnack curve gives a two-sheeted cover
of the interior of its amoeba, so the Riemann surface of the curve is easy to visualize.
In particular, the boundary of each hole lifts to a cycle on this Riemann surface.
All such cycles are independent, and it is the holonomy around these cycles and their
duals that parametrizes the bundle $M_x$.  This gives us the count of moduli. If the
number of the integer internal points of the Newton polygon $N_x$ is ${\rm Int}\, N_x$, 
then as we argue presently, the curves $\Sigma_x$ are parameterized by
${\rm Int}\,  N_x$ complex parameters.  On the other hand, when $\Sigma_x$ is a Harnack
curve, counting the moduli of the line bundle $M_x$ is particularly convenient; and as
we have just argued, $M_x$ also depends on ${\rm Int}\, N_x$ complex parameters.
The conclusion is that the moduli space has $4\times {\rm Int}\, N_x$ real dimensions. 

In fact, one does not have to work at the point where the curve saturates the bound
(\ref{AmoebaIneq}) as we did above.  Owing to a theorem of Khovanskii \cite{Khovanskii},
the genus $g$ of $\Sigma_x$ equals ${\rm Int}\, N_x$; while the number of punctures is
equal to $p=r_{-0}+r_{+0}+\sum_{\pm,j}r_{\pm j}$, the number of integer points on the boundary of $N_x$.  A holomorphic line bundle over
a Riemann surface $\Sigma_x$ is equivalent to a flat connection on a U(1) bundle over
$\Sigma_x$.  The latter is determined by its holonomy around the generators of
$\pi_1(\Sigma_x)$.  The monodromy around the punctures is fixed by the asymptotic
conditions $q_l$, and thus the remaining parameters are the $2 g$ holonomies around
the generating cycles.

This counting gives exactly the same answer for the $y$-spectral data, since  $N_x=N_y$
as we now argue. The tentacles of an amoeba exponentially approach straight lines,
and are orthogonal to the edges of the Newton polygon \cite{GKZ}.  From the construction
of the amoeba $A_x$ it is clear that the tentacles are determined by the asymptotic
eigenvalues of the Higgs field.  It follows that $A_y$ has the same asymptotes as
$A_x$, and therefore
\begin{equation}
N_x=N_y.
\end{equation}

Consider an edge $r_{-j}e_{-j}$ of a Newton polygon directed along $(-\alpha, \beta)$; then the
associated edge polynomial has the form $G_{r_{-j}e_{-j}}(s,t)=s^{k_1} t^{k_2} R(s^{-\alpha} t^\beta),$ where $(k_1,k_2)$ is the tail of the edge $r_{-j}e_{-j}$ of the Newton polygon and $R$ is a degree $r_{-j}$ polynomial.
The corresponding asymptotes of its amoeba satisfy
\begin{equation}\label{tent}
-\alpha\log s+\beta\log t=\text{const.}
\end{equation}
It follows that the associated charge is $Q=\alpha/\beta$, and that the constants appearing on the right-hand-side of \eqref{tent} are the roots of the polynomial $R.$

Let us compare this with the asymptotic conditions \eqref{BC}.  Since $t_{\pm,l}$ is an eigenvalue of the holonomy of the $D_x+\ii\Phi$ operator, it follows that the leading behaviour of the corresponding sheet of the spectral curve is
\begin{equation}
t_{\pm,l}=s^{Q_{\pm,l}} e^{2\pi(M_{\pm,l}+i p_{\pm,l})}.
\end{equation}
We conclude that for the edge $r_{-j}e_{-j}$ its edge polynomial is  $G_{r_{-j}e_{-j}}(s,t)=s^{k_1} t^{k_2} R(s^{-\alpha} t^\beta),$ with $R$ a polynomial of degree $r_{-j}$ and roots  $m_{-j}^\nu=\exp(2\pi \beta(M_{-j}^\nu+i p_{-j}^\nu)),$ $\nu=1,\ldots,r_{-j}.$
A similar comparison for an edge along $e_{+j}=(\alpha, -\beta)$ leads to $G_{e_{+j}}=s^{k_1} t^{k_2} R(s^{\alpha} t^{-\beta}),$ with $(k_1,k_2)$ being the tail of $e_{+j}$ and $R$ a polynomial of degree $r_{+j}$ and roots $m_{+j}^\nu=\exp(-2\pi \beta(M_{+j}^\nu+i p_{+j}^\nu)),$ $\nu=1,\ldots,r_{+j}.$

For the southern edge $r_{-0}e_{-0}$ the polynomial $G_{r_{-0}e_{-0}}=s^k R(\frac{1}{s}),$ with $R$ a degree $r_{-0}$ polynomial with roots $m_{-0}^\nu=\exp(-2\pi(z_{-,\nu}-\ii y_{-,\nu})).$  For the northern edge $r_{+0}e_{+0},$ on the other hand, $G_{r_{+0}e_{+0}}=s^k R(s),$ with $R$ of degree $r_{+0}$ and with roots $m_{+0}^\nu=\exp(2\pi(z_{+,\nu}-\ii y_{+,\nu})).$

So far, we have demonstrated that the Newton polygon $N_x=N_y$ determines the charges $Q$,
and moreover that it can be reconstructed from the charges and their multiplicities.  We also demonstrated that the boundary and singularity conditions of the monopole wall determine (up to an overall scaling) the coefficients of the Newton polynomial that correspond to the integer points lying on the boundary of the Newton polygon.

We would like to decorate the Newton polygon, so that a decorated Newton polygon $N_x$
is equivalent to the set of boundary data $Q, M, p, q,\vr.$  
The perimeter of a Newton polygon is divided into subintervals by all integer points on it. 
\begin{itemize}
\item A horizontal North or South subinterval is associated to, respectively, a positive
or a negative singularity.  Let us mark the value
$m_{\pm 0}^{\nu}=\exp(\pm2\pi(z_{\pm,\nu}-\ii y_{\pm,\nu}))$ and
$\tilde{m}_{\pm 0}^{\nu}=\exp(\pm2\pi(z_{\pm,\nu}+\ii x_{\pm,\nu}))$ next to each
subinterval of respectively $N_x$ and $N_y$.
Here $\vr_{\pm,\nu}=(x_{\pm,\nu},y_{\pm,\nu},z_{\pm,\nu})$ are the positions of
the positive and negative singularities. 
\item Any other subinterval, however, is associated with an asymptotic eigenvalue
labelled by $(Q_{\pm j},M_{\pm j}^\nu)$; to be exact, for $Q=\alpha/\beta$
with $\alpha$ an integer and $\beta$ a positive integer, $\alpha$ and $\beta$ being
relatively prime. Such a subinterval with
$Q=\alpha_{\pm j}/\beta_{\pm j}=\alpha_{\pm, l}/\beta_{\pm,l}$ corresponds
to $\beta_{\pm j}=\beta_{\pm,l}$ degenerate eigenvalues corresponding to
$(Q_{\pm, l},M_{\pm, l})$.  We mark
$m_{\pm j}^\nu=\exp(\mp2\pi\beta_{\pm j}(M_{\pm j}^\nu+i p_{\pm j}^\nu))=m_{\pm,l}=\exp(\mp2\pi\beta_{\pm,l}(M_{\pm,l}+i p_{\pm,l}))$ next to that subinterval
of $N_x$ and $\tilde{m}_{\pm,l}=\exp(\mp2\pi\beta(M_{\pm,l}-i q_{\pm,l}))$
next to that subinterval of $N_y$.
\end{itemize}

As one moves along the perimeter of the Newton polygon, applying Vieta's theorem to each edge, one finds that the values $m_{\pm j}$ have to satisfy $\prod_{j,\nu} (-m_{\pm j})=1.$ This is the reason for the relations \eqref{Vieta}.   Of course, the y-spectral data via the same reasoning lead to $\prod_{j,\nu} (-\tilde{m}_{\pm j})=1.$  In more detail, for any given perimeter edge $e_j$ its polynomial $G_{e_j}$ has the product $\prod_{\nu}(-m_j^\nu)$ of the negatives of its roots equal to the ratio of its head to tail term coefficients.  Since all perimeter edges form a closed loop, the product of their head to tail coefficient ratios equals to one.  Thus
\begin{align}
&\sum_{\nu=1}^{r_{+0}} z_{+,\nu}-\sum_{\nu=1}^{r_{-0}} z_{-,\nu}-\sum_{j,\nu} \beta_{+ j} M_{+j}^\nu+\sum_{j,\nu} \beta_{- j} M_{-j}^\nu=0,\\
\label{FirstDir}
&\sum_{\nu=1}^{r_{+0}} y_{+,\nu}-\sum_{\nu=1}^{r_{-0}} y_{-,\nu}+\sum_{j,\nu} \beta_{+ j} p_{+j}^\nu-\sum_{j,\nu} \beta_{- j} p_{-j}^\nu\in\mathbb{Z}+\frac{1}{2}\sum_{\pm,j}r_{\pm j},\\
\label{SecondDir}
&\sum_{\nu=1}^{r_{+0}} x_{+,\nu}-\sum_{\nu=1}^{r_{-0}} x_{-,\nu}+\sum_{j,\nu} \beta_{+ j} q_{+j}^\nu-\sum_{j,\nu} \beta_{- j} q_{-j}^\nu\in\mathbb{Z}+\frac{1}{2}\sum_{\pm,j}r_{\pm j},
\end{align}
Since $M_{\pm j}^\nu$ appears $\beta_{\pm j}$ times among $M_{\pm,l},$ the above relations give rise to Eqs.~\eqref{Vieta} once we evaluate the parity of the number of perimeter points.  As we demonstrate momentarily (Eq.~\eqref{Eq:DimQ}), the
 shift $\frac{1}{2}\sum_{\pm,j}r_{\pm j}=\frac{1}{2}p$ on the right-hand-side of relations \eqref{FirstDir} and \eqref{SecondDir} can easily be computed in terms of the charges:
\begin{equation}
\frac{1}{2}\sum_{\pm,j}r_{\pm j}\equiv\frac{1}{2}\sum_{\stackrel{l_1,l_2=1} {\mbox{\tiny $l_1<l_2$}}}^{2n} (Q_{,l_1}-Q_{,l_2}) \mod {\mathbb{Z}}.
\end{equation}

The pair of decorated Newton polygons completely determines the boundary conditions:
the asymptotics and the singularities.  The spectral polynomial, on the other hand,
has its perimeter terms determined by the markings.  The only free parameters in
determining the spectral curve are the coefficients of the terms corresponding
to the internal points of $N$.  
Thus the real dimension of the moduli space equals four times the number of
integer points strictly inside the Newton polygon:
\begin{align}
{\rm dim}\, {\cal M}&=4\, {\rm Int}\, N_x,
\end{align}
Now we would like to compute this dimension in terms of the asymptotic data.  
Let us assemble the elementary vectors of \eqref{ElementaryVec} as follows:
\begin{equation}
(e_{\bf j})=(e_{-0}, e_{-1},\ldots, e_{-f_-}, e_{+0}, e_{+1},\ldots, e_{+f_+});
\end{equation}
and whenever the edge $e_{\bf j}$ is not horizontal, let $Q_{\bf j}$ denote the
charge corresponding to $e_{\bf j}$ if ${\bf j}\neq\pm0$. 
The area of the Newton polygon can be computed from the determinant formula
applied to its sequence of edges:
\begin{equation}
(r_{\bf j} e_{\bf j})=(r_{-0}e_{-0}, r_{-1}e_{-1},\ldots,
  r_{-f_-}e_{-f_-}, r_{+0}e_{+0}, r_{+1}e_{+1},\ldots, r_{+f_+}e_{+f_+}):
\end{equation}
\begin{equation}
A(N)=\frac{1}{2}\left|\sum_{\stackrel{{\bf i},{\bf j}=1}
  {\mbox{\tiny ${\bf i}<{\bf j}$}}}^{f_-+f_++2} r_{\bf i} r_{\bf j}
    e_{\bf i}\times e_{\bf j}\right|.
\end{equation}  
Note that if $e_{\bf i}$ and $e_{\bf j}$ are not horizontal, then
$r_{\bf i}\, r_{\bf j}\, e_{\bf i}\times e_{\bf j}=
(r_{\bf i}\beta_{\bf i})(r_{\bf j}\beta_{\bf j})(Q_{\bf i}-Q_{\bf j})$,
and $r_{\bf i}\beta_{\bf i}$ is a multiplicity of an eigenvalue with charge
$Q_{\bf j}$.  On the other hand, for the horizontal $e_{\bf i}$ the contributions
to the sum are 
$r_{-0}\,r_{\pm j}\,e_{-0}\times e_{\pm j}=\pm r_{-0}(r_{\pm j}\beta_{\pm j})$, 
$r_{-j}\,r_{+0}\,e_{-j}\times e_{+0}=- r_{+0}(r_{-j}\beta_{- j})$ and
$r_{+0}\,r_{+ j}\,e_{+0}\times e_{+ j}=- r_{+0}(r_{+ j}\beta_{+ j}).$

The number $p$ of integer points on its perimeter is given by the sum of
multiplicities $p=\sum_{\bf j} r_{\bf j}$.  Now Pick's formula for the
area allows us to find the number of integer internal points:
\begin{equation}
{\rm Int}\, N=A(N)-\frac{p}{2}+1.
\end{equation}
This gives the dimension of the moduli space 
\begin{align}\label{Eq:DimQ}
{\rm dim}\, {\cal M}&=\left|n r_{+0}-\frac{1}{2}\sum_{\stackrel{l_1,l_2=1}
  {\mbox{\tiny $l_1<l_2$}}}^{2n} (Q_{,l_1}-Q_{,l_2})\right|-\frac{p}{2}+1.
\end{align}
Here the set of charges is
$(Q_{,l})=(Q_{-,1},Q_{-,2},\ldots,Q_{-,n}, Q_{+,1}, Q_{+,2},\ldots, Q_{+,n})$,
with $Q_{-,1}\geq Q_{-,2}\geq \ldots\geq Q_{-,n}$ and
$ Q_{+,1}\geq Q_{+,2}\geq\ldots\geq Q_{+,n}$,  so the index $l$ ranges
from $1$ to $2n$.  We note that in the sets $Q_-$ and $Q_+$ of Eqs.~\eqref{Eq:Qs1}
and \eqref{Eq:Qs2}, each $Q_{\pm j}$ appears only once, while in the above
$2n$-plet $(Q_{,l})$ this value of the charge appears $r_{\bf j}\beta_{\bf j}$
times, so that each eigenvalue of the Higgs field contributes one term to it.

\section{Nahm Transform and $SL(2,\mathbb{Z})$ Action}\label{Sec:Nahm}
\subsection{The Nahm transform}\label{Sec:NTransf}
The generalized Nahm transform \cite{Nahm82} maps a doubly-periodic monopole to another
doubly-periodic monopole. Our conventions for the transform are as follows.
The dual coordinates are denoted $(x',y',z')=(x'^1,x'^2,x'^3)$, with $x'$ being
dual to $y$, and $y'$ being dual to $x$, and with $x',y'$ each having period~1.
Given a monopole field, one looks for normalizable solutions of $\Delta\Psi=0$, where
\begin{equation} \label{Delta}
  \Delta = \left[\begin{array}{cc}
            D_z-\ii\Phi+2\pi z' & D_x-\ii D_y+2\pi\ii(y'-\ii x')\\
            D_x+\ii D_y+2\pi\ii(y'+\ii x') & -D_z-\ii\Phi+2\pi z'
         \end{array}\right].
\end{equation}
Let us denote the dimension of the space of such solutions by $n'.$
The $n'$ orthonormalized solutions are assembled as the columns of $\Psi$, and
one then defines the Nahm-transformed fields $A'=A'_jdx'^{j}$ and $\Phi'$,
which have rank $n'$, by
\begin{eqnarray} \label{NahmTransform}
  A'_j &=& \int\Psi^{\dagger}\frac{\pa}{\pa(x')^j}\Psi \,d^3x, \\
  \Phi' &=& -2\pi\ii\int z\,\Psi^{\dagger}\Psi \,d^3x.
\end{eqnarray}
The spectral curve $\widetilde{\Sigma}_{x'}$ of the Nahm-transformed field lies
in the same space as $\Sigma_x$, and the two curves are in fact identical.
More precisely, $V_{x'}(t)$ is defined by integrating
$(D_{x'}+\ii\Phi')\psi=0$ in the $x'$-direction, where
$t=\exp[2\pi(z'-\ii y')]$; and the curve $\widetilde{\Sigma}_{x'}$
defined by $F_{x'}(t,s):=\det[V_{x'}(t)-s]=0$ is identical to $\Sigma_x$.
In other words, $F_{x'}(t,s)$ and $F_{x}(s,t)$ have the same zeros;
in effect, the spectral data of the Nahm-transformed field is
obtained by interchanging $s$ and $t$.
The $y$-spectral curve $\Sigma_y$ is similarly invariant under
the Nahm transform.

\subsection{$SL(2,\mathbb{Z})$ Action}
We mentioned that $(\Sigma_x, M_x)$ uniquely determines a monopole wall up
to gauge transformations.  The spectral curve embedding
$\Sigma_x\subset\mathbb{C}^*_s\times\mathbb{C}^*_t$ is central in
reconstructing the solution.  Let us consider an $SL(2,\mathbb{Z})$ change
of coordinates
\begin{equation}\label{SL2Z}
g=\left(\begin{matrix} a & b\\c & d \end{matrix}\right): (s,t)\mapsto (s^d t^c, s^b t^a),
\end{equation}
with $ad-bc=1$. This induces a map on the pairs $(\Sigma_x, M_x)$, and thus
on monopole-wall solutions. 
The $SL(2,\mathbb{Z})$ action above is chosen so that under this action we have
$s^{-\alpha}t^\beta\mapsto (s')^{-\alpha'}(t')^{\beta'}$ with
$\left(\begin{smallmatrix}\alpha'\\ \beta'\end{smallmatrix}\right)=
\left(\begin{smallmatrix} a&b\\ c&d\end{smallmatrix}\right)
\left(\begin{smallmatrix}\alpha\\ \beta\end{smallmatrix}\right)$.
Thus in the plane of the Newton polygon, $g$ acts via
$\left(\begin{smallmatrix} a& -b\\ -c&d\end{smallmatrix}\right)$, and the resulting
Newton polygon $N'=\left(\begin{smallmatrix} a& -b\\ -c&d\end{smallmatrix}\right)N$
is the image under this linear map.  Moreover, it is decorated so that the same
label $m$ is associated to a subedge of $N$ and to its image subedge $e'$ of $N'$.
This determines the resulting boundary conditions, the singularities, and the rank of
the $g$-transformed monopole wall.  Let us now write these out explicitly.

For a U($n$) monopole wall with $r_{-0}$ negative and $r_{+0}$ positive
singularities and charges $Q_{\pm j}=\alpha_{\pm j}/\beta_{\pm j}$,
we would like to know how these quantities transform under the $SL(2,\mathbb{Z})$
action above.  Let us distinguish the corresponding quantities for the
$SL(2,\mathbb{Z})$ transform of the original monopole wall by a prime.
Under the action of an element $g$ of \eqref{SL2Z}, the rank goes to
$n'=\frac{1}{2}\sum_{\bf j} r_{\bf j} |c\alpha_{\bf j}+d\beta_{\bf j}|$. 

Altogether, the set of vector-multiplicity pairs
$\{ (e'_{{\bf j}'}, r_{{\bf j}'}) \}=
\left\{ \left(\left(\begin{smallmatrix} a & -b\\ -c & d \end{smallmatrix}\right) 
e_{\bf j}, r_{\bf j} \right) \right\}$ gives the sequence of edges
$r_{{\bf j}'} e'_{{\bf j}'}$ of $N'$  each along an elementary vector
$e'_{{\bf j}'}$. The vectors $e'_{{\bf j}'}$ in this set which have the
form $\left(\begin{matrix} \alpha_{+j'} \\ -\beta_{+j'}\end{matrix}\right)$,
{\sl i.e.}\ with negative second component, correspond to  the charges
$Q'_{+j'}=\alpha_{+j'}/\beta_{+j'}$. The vectors $e'_{{\bf j}'}$ of the
form $\left(\begin{matrix} -\alpha_{-j'} \\ \beta_{-j'}\end{matrix}\right)$,
{\sl i.e.}\ with positive second component, correspond to  the charges
$Q'_{-j'}=\alpha_{-j'}/\beta_{-j'}$.
The constant terms, on the other hand, transform as 
$M_-=\big\{ \frac{M_j}{c Q_j+d}\, \big|\, c\alpha_j+d\beta_j>0\big\}$ and 
$M_+=\big\{ \frac{M_j}{c Q_j+d}\, \big|\, c\alpha_j+d\beta_j<0\big\}$;
and similarly for $p_\pm$ and $q_\pm$.  Perhaps a simpler way of formulating these rules is 
stating that the asymptotic corresponding to some subedge $e_{\bf j}=(e',e'')$ satisfies 
$(s,t)^{e_{\bf j}}\equiv s^{e'}t^{e''}=m_{\bf j}^\nu$ and the constants $m_{\bf j}^\nu$
are $SL(2,\mathbb{Z})$ invariant.

The Nahm transformation of section \ref{Sec:NTransf} is identified with
$\left(\begin{smallmatrix} 0 &1\\1&0 \end{smallmatrix}\right)$, which is not
an element of $SL(2,\mathbb{Z})$; however, a reflection of the $z$ and $y$
coordinates followed by the Nahm transform corresponds to the element
$S=\left(\begin{smallmatrix} 0 & -1\\1&0 \end{smallmatrix}\right)$  of the modular
group\footnote{In fact, of course, reflections extend the action of $SL(2,\mathbb{Z})$ to the action of the full general linear group $GL(2,\mathbb{Z})$ group on monopole walls.}.  The action of the second generator
$T=\left(\begin{smallmatrix} 1 & 1\\0&1 \end{smallmatrix}\right)$ on monopole
walls is more prosaic.  For any U($n$) solution $(A,\Phi)$ and any U(1) solution
$(a_a,\phi_a),$ their sum $(A+\ii a_a\mathbb{I}, \Phi+\ii \phi\mathbb{I})$
is another solution.  What is the influence of this operation on the spectral
curve $\Sigma_x: \{F_x^A(s,t)=0\}$ of the first solution?  If the spectral
curve of $(a_a,\phi_a)$ is given by $t=P(s)/Q(s)$, then the holonomy is
$V_x^{A+a_a}=V_x^A P(s)/Q(s)$, and therefore $s'=s$ and
\begin{multline}
F_x^{A+a}(s',t')={\rm det}\left[V_x^{A+a_a}-t'\right]\\
=\left(\frac{P(s)}{Q(s)}\right)^n{\rm det}\left[V_x^{A}-\frac{Q(s)}{P(s)}t'\right]=\left(\frac{P(s)}{Q(s)}\right)^nF_x^A\left(s,\frac{Q(s)}{P(s)}t'\right).
\end{multline}
Thus the transformation $T=\left(\begin{smallmatrix} 1 &1\\ 0&1 \end{smallmatrix}\right)$
is identified with the addition of the simplest constant-energy abelian
solution \eqref{ConstE} of charge 1 with $P(s)/Q(s)=s$.


\section{Perturbative Approach}

In this section, we take $\Phi$ and $A_j$ to be su(2)-valued and smooth (no Dirac
singularities). The aim is to look at two examples: one having no moduli,
and the second having four moduli. In the second case, there is an explicit
solution corresponding to one particular point in the moduli space, and we
investigate the tangent space at that point by solving the Bogomolmy equation
linearized about that solution. Since the explicit solution is highly symmetric,
the calculation is a delicate one.

There are two topological charges $Q_{\pm}\in\ZZ$, and the set of 
asymptotic parameters  $\{ M_{\pm}, p_{\pm}, q_{\pm} \}$.
The spectral function $F(s,t)=\det[V_x(s)-t]$ takes the form
$F(s,t)=t^2-W_x(s)t+1$, where $W_x(s)=\tr V_x(s)$. Similarly,
integrating in the $y$-direction gives a function $W_y(\st)=\tr V_y(\st)$. The
$x,y$-holonomy parameters $(p_{\pm},q_{\pm})$
show up in the asymptotic behaviour of $W_x$ and $W_y$.
For example, in the $(Q_{-}, Q_{+})=(0,1)$ case described below, we have
\begin{eqnarray} 
   W_x(s) &=& 2\cosh[2\pi(M_- + \ii p_-)] + 
                s\,\exp[2\pi(M_+ + \ii p_+)], \label{Wfor10caseA}\\
   W_y(\st) &=& 2\cosh[2\pi(M_- + \ii q_-)] + 
               \st\,\exp[2\pi(M_+ + \ii q_+)], \label{Wfor10caseB}
\end{eqnarray}
and these expressions define $(p_{\pm},q_{\pm})$.

For $Q_{\pm}\geq0$, one expects the existence of monopole-wall solutions
containing $N=Q_{+}+Q_{-}$ monopoles per unit cell. The centre-of-mass
of these monopoles is determined by the asymptotic parameters, leaving
$4(N-1)$ moduli. The functions $W_x(s)$ and $W_y(\st)$ will have the form
\begin{eqnarray} 
   W_x(s) &=& A_0\, s^{Q_{+}} + \ldots + A_N s^{-Q_{-}},
                \label{WforgeneralSU2caseA}\\
   W_y(\st) &=& \widetilde{A}_0\, \st^{Q_{+}} + \ldots
                 + \widetilde{A}_N \st^{-Q_{-}};\label{WforgeneralSU2caseB}
\end{eqnarray}
the ``external'' coefficients $\{A_0,A_N,\widetilde{A}_0,\widetilde{A}_N\}$
are completely determined by the asymptotic data, whereas the ``internal'' coefficients
$\{A_1,\ldots,A_{N-1},\widetilde{A}_1,\ldots,\widetilde{A}_{N-1}\}$
are the moduli.


\subsection{The case $(Q_{-}, Q_{+})=(0,1)$}
It was remarked earlier that the Nahm transform of a
centred Dirac 2-pole wall is an SU(2) wall. This SU(2) solution
is the one having charges $(0,1)$, and it will be described in this subsection.
Recall that a centred Dirac 2-pole wall depends on six parameters:
the pole position $\vb$, modulo $\vb\mapsto-\vb$ which
corresponds to interchanging the two poles; and the U(1) asymptotic data
$(M',p',q')$, where a tilde is used to distinguish
these parameters from the SU(2) asymptotic data. The U(1) asymptotic data
as $z\to-\infty$ are equal to the $z\to+\infty$ data, because of the centring,
so $\pm$ subscripts are omitted on these. Then comparing
(\ref{Wfor10caseA}, \ref{Wfor10caseB})
and the formulae from the end of section 3.1.4 shows that
\[
  M_+=-M',\quad r_+=-(p'+\ii q'),
   \quad M_-=b_3,\quad r_-=\ii(b_1+\ii b_2),
\]
where $r_{\pm}=p_{\pm}+\ii q_{\pm}$. Note that the field with data
$(-M_-,-r_-)$ is gauge-equivalent to the one with data
$(M_-,r_-)$: this corresponds to the interchange $\vb\mapsto-\vb$. So the
parameter space is $(T^2\times \RR)\times(T^2\times \RR)/\ZZ_2$.

The solution resembles a wall of smooth SU(2) monopoles, with one monopole
per unit cell. This monopole is located at
\[
   (x_0, y_0, z_0)\approx(\half+q_--q_+, \half-p_-+p_+, M_--M_+).
\]
It separates a vacuum region (for $z<z_0$) from a region of constant energy
density (for $z>z_0$). The size of the monopole, relative to the $x,y$-period,
is determined by $\mu = 2\pi M_- = \lim_{z\to-\infty} |\Phi|$. The solution
is completely determined by its asymptotic data: there are no moduli.

\begin{figure}[hbt] 
  \begin{center}
  \includegraphics[scale=0.5]{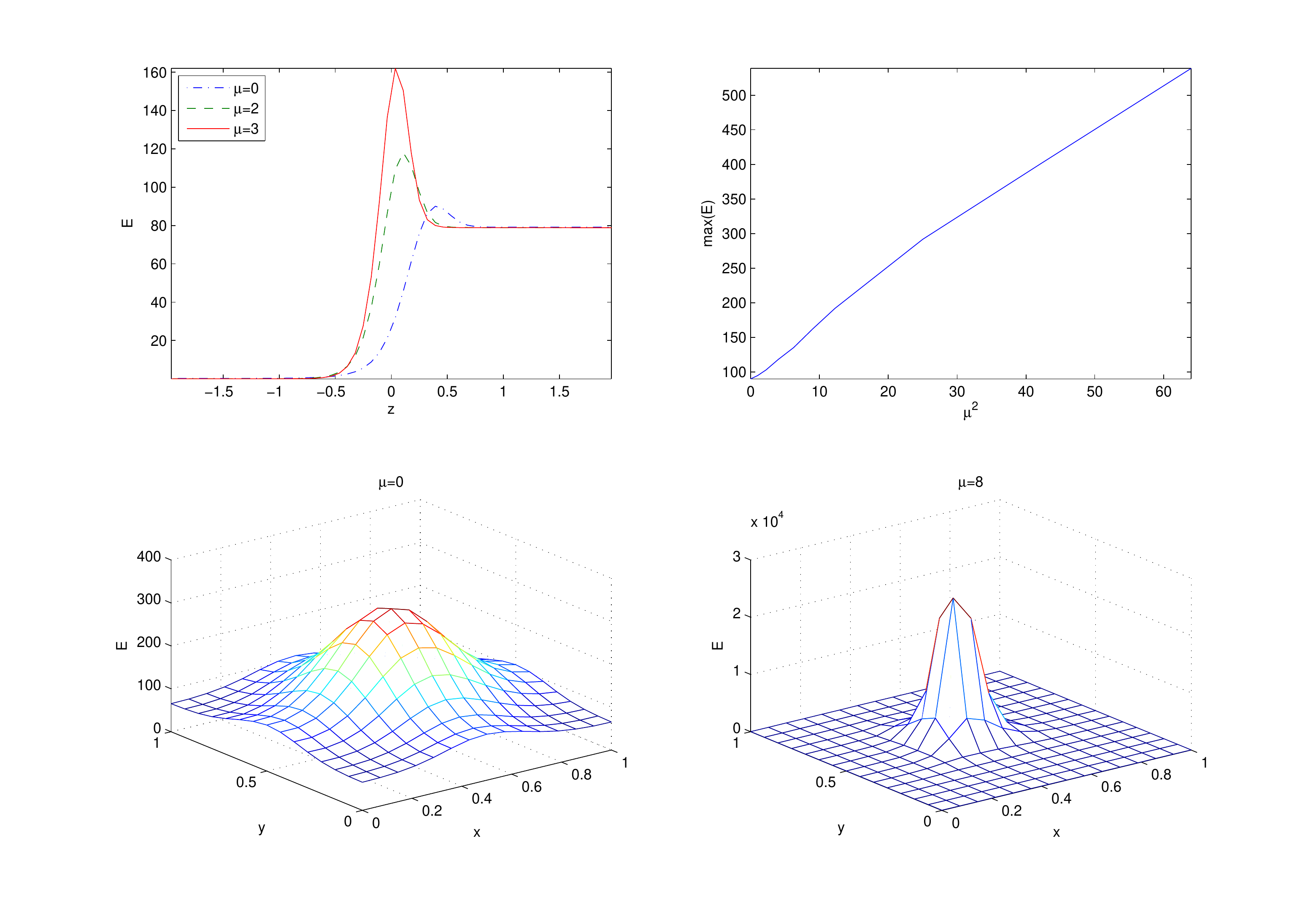}
  \caption{Properties of the {\rm SU(2)} 1-monopole wall.\label{1monwall}}
  \end{center}
\end{figure}
Numerical investigation, using the numerical lattice-gauge-theory method
described in \cite{Ward:2006wt}, gives the picture illustrated in
Figure~\ref{1monwall}. In these examples, we take
$p_{\pm}=q_{\pm}=M_--M_+=0$, so we get a monopole
centred at $(x_0, y_0, z_0)\approx(\half, \half, 0)$, with various values of
$\mu$. The upper left-hand plot shows the
energy density $E$, integrated over one $x,y$-cell, as a function of
$z$, for three values $\mu=0,2,3$ of the parameter $\mu$; the domain-wall
nature of the solution is clearly visible. It is clear that the maximum
value of the energy density (the `height' of the wall) increases with
$\mu$: this is also illustrated in the upper right-hand subfigure, which
plots this maximum as a function of $\mu^2$.
The two lower subfigures plot the energy density,
restricted to $z=z_0$, as a function of $x$ and $y$, for the two cases
$\mu=0$ and $\mu=8$. As expected, one sees a single monopole located
at $x=y=0.5$. Note that the Higgs field $\Phi$ has a single zero which
coincides, at least approximately, with the energy maximum in this case.
The figures also illustrate the fact that as $\mu$ increases,
the monopoles in the wall become more localized ({\sl i.e.}\ smaller
relative to the $xy$-period).

Finally, let us consider the $z$-spectral data.
As before, we centre the field in space by taking $M_+=M_-=M>0$,
$p_{+}=p_{-}=p$, and $q_{+}=q_{-}=q$. So the remaining asymptotic data
are $M$ and $r=p+\ii q$. In order to compute the $z$-scattering function $B$,
we need to choose solutions $\psi$ and $\psi_{-}$ as described in section 3.1.3.
Let us say that the standard gauge as $z\to\infty$ is one in which
\begin{equation}\label{01plusinfty}
   \Phi=2\pi\ii(z+M)\sigma_3 + O(z^{-2}), \quad 
           A_j=\pi\ii(y-2p,-x-2q,0)\sigma_3 + O(z^{-1})\quad\mbox{as $z\to\infty$};
\end{equation}
and the standard gauge as $z\to-\infty$ is one in which
\begin{equation}\label{01minusinfty}
   \Phi=-2\pi\ii M\sigma_3 + O(z^{-2}), \quad 
           A_j=2\pi\ii(p,q,0)\sigma_3 + O(z^{-1})\quad\mbox{as $z\to-\infty$}.
\end{equation}
%
The remaining gauge freedom for large $|z|$ consists of constant diagonal
gauge transformations. It is then consistent to choose $\psi$ and $\psi_{-}$
such that
\begin{eqnarray}
  \psi\,\exp\{\pi(z+M)^2\} &\to& \left[\begin{array}{c} 0 \\
         \exp\{-\ii\pi\zeta(y+r)-2\pi ry\}\end{array}\right]
           \mbox{ as $z\to\infty$},\label{01psi}\\
  \psi_{-}\,\exp\{2\pi Mz\} &\to& \left[\begin{array}{c} \exp(-\ii\pi r\bar{\zeta}) \\
             0\end{array}\right] \mbox{ as $z\to-\infty$}\label{01psiminus}
\end{eqnarray}
in the relevant standard gauge, where $\zeta=x+\ii y$. These solutions are indeed
covariantly holomorphic in $\zeta$. The relation (\ref{01psi}) defines $\psi$
uniquely, and (\ref{01psiminus}) defines $\psi_{-}$ up to adding a solution
vanishing as $z\to-\infty$ (which does not affect $B$). It then follows
from this particular choice that $B(\zeta)$ has the periodicity
behaviour $B|_{x=1}=B|_{x=0}$ and
$B|_{y=1}=\exp(\pi-2\ii\pi\zeta)\,B|_{y=0}$; and this implies that
$B$ is a theta-function, namely
\begin{equation}\label{Bscattering}
  B(\zeta) = c\,\vartheta_3(\pi\zeta)
\end{equation}
for some constant $c$. (The phase of $c$ is undetermined because of the remaining
constant gauge freedom, while the modulus of $c$ depends on the normalization of the
basis vectors.)
In particular, there is exactly one spectral line in this case.
The corresponding numerical solutions have the feature that the monopole
is located, to within the numerical accuracy, on this spectral line.


\subsection{The case $Q_{\pm}=1$}
An explicit example with $Q_{\pm}=1$ is the constant-energy solution, obtained
by taking
\begin{equation} \label{const-energy}
  \Phi=2\ii\pi z\sigma_3, \qquad A_j=(\ii\pi y, -\ii\pi x, 0)\sigma_3,
\end{equation}
and then making a (non-periodic) SU(2) gauge transformation so that the fields
$(\Phi, A_j)$ become periodic in $x$ and $y$. For this solution, the energy
density has the constant value ${\cal E}=8\pi^2$: hence its name.
The prescription above, involving as it does a
non-periodic gauge transformation, does not fully determine the field; in
particular, it does not fix the holonomy of the gauge field in the $x$ and
$y$ directions. This ambiguity can be removed, without loss of generality,
by saying that the holonomy is chosen so that the functions
$W_x$ and $W_y$ are given by
\begin{equation} \label{const-energy-W}
  W_x(s)=s+s^{-1}, \qquad W_y(\st)=\st+\st^{-1}.
\end{equation}
Note that the equation for the spectral curve $\Sigma_x$ can be written as
$t+t^{-1} = s+s^{-1}$, from which it is clear that $\Sigma_x$ is invariant
under the interchange $s\leftrightarrow t$. This is consistent with the
fact that this constant-energy solution is invariant under the Nahm
transform.

This solution belongs to a family containing several parameters and moduli.
The former are the asymptotic data $(M_{\pm},p_{\pm},q_{\pm})$; these show
up in the spectral functions $W_x(s)$ and $W_y(\st)$, which have the general form
\begin{equation} \label{const-energy-W-deformed}
  W_x(s)=D_+s+D+D_-s^{-1},
     \qquad W_y(\st)=\widetilde{D}_+\st+\widetilde{D}+\widetilde{D}_-\st^{-1},
\end{equation}
with $D_{\pm}=\exp[\pm2\pi(M_{\pm}+\ii p_{\pm})]$ and
$\widetilde{D}_{\pm}=\exp[\pm2\pi(M_{\pm}+\ii q_{\pm})]$. In effect,
the asymptotic data include position parameters. For example, $M_++M_-$ is
translation-invariant; but $M_+-M_-$ is changed by a $z$-translation, and
corresponds to the location of the wall.
\begin{figure}[hbt] 
  \begin{center}
  \includegraphics[scale=0.5]{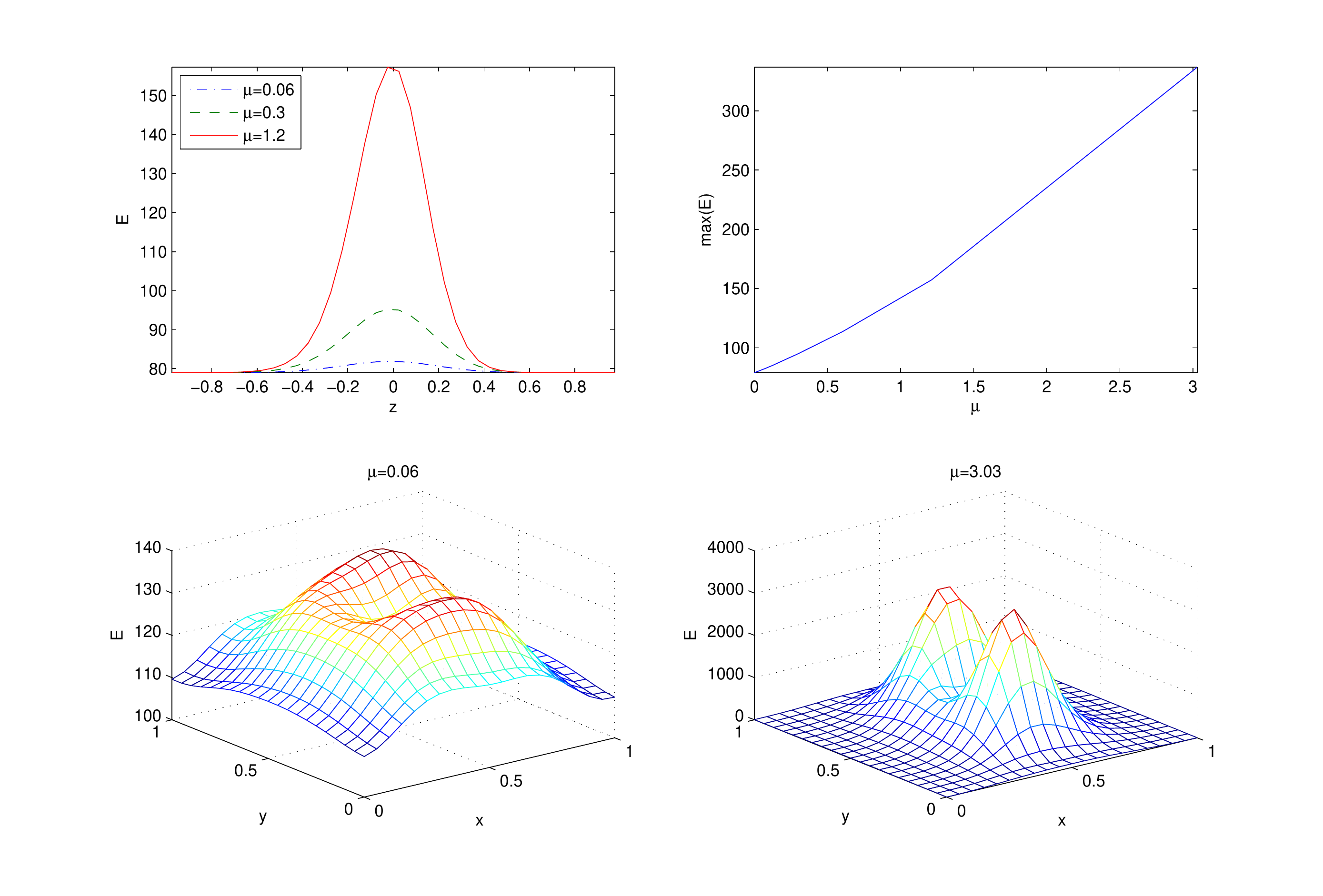}
  \caption{A family of  $Q_{\pm}=1$ walls.\label{Qpm1walls}}
  \end{center}
\end{figure}
The effect of varying the asymptotic parameter $\mu = 2\pi M_+ = 2\pi M_-$
is illustrated in Figure~\ref{Qpm1walls}. This shows a family of
numerically-generated solutions, the subfigures being the analogues of
those in Figure~\ref{1monwall}.
For $\mu=0$ we have the constant-energy solution; but as $\mu$ increases,
one gets a wall of increasing height and spatial localization, with two
monopoles in each cell as expected.
In each case, the wall is located at $z=0$; the two lower subfigures
show the energy density restricted to $z=0$, for two different values of $\mu$.

For the remainder of this subsection,
we restrict to fields which have the same asymptotics as those of
the constant-energy field, so $D_{\pm}=\widetilde{D}_{\pm}=1$.
There are four moduli, which determine the ``interior''
constants $D$ and $\widetilde{D}$. The corresponding
4-dimensional space of infinitesimal perturbations can,
in fact, be obtained by explicitly solving the Bogomolny equation
linearized about the constant-energy solution, restricting to
perturbations which are localized in $z$. More precisely, the
space of first-order perturbations is 8-dimensional, but only a
4-dimensional subspace of these can be extended to second order ---
the generic first-order perturbation becomes non-local in $z$ at
second order, and changes the asymptotics ($M_++M_-$ etc).

A sketch of the details is as follows. At first order, the calculation is
analogous to those in \cite{Lee:1998isa, Ward:2005nn}. One adds perturbations
$\vep\phi=\ii\vep\phi^a\sigma_a$ and $\vep a_j=\ii\vep a^a_j\sigma_a$
to the constant-energy field (\ref{const-energy}), with $\vep$ being small.
The $z$-localized solutions turn out to be given by $\phi^3=0=a^3_j$ and
\begin{eqnarray}
   a^1_1+\ii a^2_1 &=& \ii(a^1_2+\ii a^2_2)
      = f(\bar{\zeta})\exp(-2\pi z^2-2\pi\ii\bar{\zeta}y),\label{f11}\\
   a^1_3+\ii a^2_3 &=& -\ii(\phi^1+\ii\phi^2)
      = \overline{g(\zeta)}\exp(-2\pi z^2-2\pi\ii\bar{\zeta}y),\label{g11}
\end{eqnarray}
where $f(\bar{\zeta})$ and $g(\zeta)$ are
Jacobi modular functions of weight  (1,2) and theta-period i.  A basis for such
functions is provided by $[\vartheta_1(\pi\zeta)]^2$ and $[\vartheta_3(\pi\zeta)]^2$.
In explicit terms, using standard theta-function conventions \cite{NIST}
(see Appendix B), $f$ and $g$ are given by
\begin{equation}
 \overline{f(\bar{\zeta})}=\overline{C}_1[\vartheta_3(\pi\zeta)]^2 +
      \overline{C}_2[\vartheta_1(\pi\zeta)]^2,
  \quad g(\zeta)=C_3[\vartheta_3(\pi\zeta)]^2+
      C_4[\vartheta_1(\pi\zeta)]^2,
\end{equation}
where the nome is $q=\ee^{-\pi}$, and $\{C_1,C_2,C_3,C_4\}$ are complex
constants.

The induced metric on the space of linear deformations gives the norm of this
perturbation, namely 
\begin{equation}\label{Eq:Metr}
2\int (f\bar{f}+g\bar{g})\,\ee^{-4\pi y^2}\ee^{-2\pi z^2}\, dx\, dy\, dz
  =\Upsilon(C_1\bar{C}_1+C_2\bar{C}_2+C_3\bar{C}_3+C_4\bar{C}_4).
\end{equation}
This calculation, and the definition of the constant $\Upsilon$, appear in
Appendix~B.

So there is an 8-real-dimensional space of first-order perturbations.
The quantities $D$ and $\widetilde{D}$ remain zero at $O(\vep)$,
but they should change at $O(\vep^2)$, which suggests extending the calculation
to second order. So we take
\[
   \Phi=2\ii\pi z\sigma_3 + \vep\phi + \vep^2\tilde{\phi},
\]
and similarly for the gauge potentials. We then solve the Bogomolny equations
to order $\vep^2$, with $(\phi,a_j)$ given by (\ref{f11}, \ref{g11}).
The $\sigma_1$- and $\sigma_2$-components of
$(\tilde{\phi},\tilde{a}_j)$ simply correspond to changing the
$C_\alpha$. The $\sigma_3$-components satisfy
\begin{equation}\label{poisson}
  \partial_j \tilde{\phi}^3 + \vep_{jkl}\,\partial_k \tilde{a}^3_l-4\,\omega_j=0,
\end{equation}
where
\[
\omega_j=\left({\rm Re}(fg),\,{\rm Im}(fg),\,|f|^2-|g|^2\right)
               \exp(-4\pi z^2 -4\pi y^2).
\]
In effect, these are Poisson equations for $(\tilde{\phi},\tilde{a}_j)$,
with a doubly-periodic source. By (say) looking at $xy$-Fourier components,
we see that they have solutions localized in $z$ if and only if the source has
no $xy$-constant term: $\int \omega_j \,dx\,dy = 0$.
This gives three real constraints on the $C_\alpha$, namely
\begin{equation}\label{Eq:Moments}
  C_1 C_3+C_2 C_4=0=|C_1|^2+|C_2|^2-|C_3|^2-|C_4|^2.
\end{equation}
In addition, there is some
residual gauge freedom, arising from the fact that (\ref{const-energy}) is invariant
under isorotation about the $\sigma_3$-axis; the action on $C_\alpha$ is to
multiply by an overall phase:
\begin{align}\label{Eq:GrAct}
C_{1,2}&\mapsto\ee^{\ii\theta}C_{1,2},&
C_{3,4}&\mapsto\ee^{-\ii\theta}C_{3,4},
\end{align}
with $\theta$ constant.
Choosing this phase gives a fourth constraint on the $C_\alpha$, and (with
a convenient choice of the phase) we get the relations $C_3=C_2$ and
$C_4=-C_1$.  According to the general Kuranishi argument (see the next
section), there are  no further obstructions at higher order, and thus there
are just two complex (or four real) moduli.  This is in perfect agreement
with our moduli count using the Newton polygon of Figure~\ref{Fig:SU2_Rombus}.

In fact, what we have computed is the moduli space metric in the vicinity of
the reducible solution \eqref{const-energy}.  In perfect accord with the general
Kuranishi theory of the next subsection, the moduli space at some solution is a
hyperk\"{a}hler reduction of the space of linear deformations by the stabilizer
group.  In our case, the space of linear deformations was flat eight
real-dimensional space, its metric being given by Eq.~\eqref{Eq:Metr}.
We may view this as a direct sum of two quaternionic lines
$\mathbb{H}\oplus\mathbb{H}$. The first factor
$\mathbb{H}\approx\mathbb{C}\oplus\mathbb{C}$ has coordinates $(C_1, C_3)$,
and the second factor has coordinates $(C_2, C_4)$. The
obstruction-vanishing condition \eqref{Eq:Moments} that ensures extension of
the linear deformation to second order is exactly the vanishing moment-map
condition for the action \eqref{Eq:GrAct} of the stabilizer group. Thus the
model for the resulting metric near the reducible solution is the hyperk\"ahler quotient 
$\mathbb{H}\oplus\mathbb{H}/\!\!/\!\!/{\rm U(1)}$, which is metrically
$\mathbb{R}^4/\mathbb{Z}_2,$ with an $A_1$ singularity at the origin.  One might worry about higher order obstructions reducing the number of deformations even further, however, the argument of the next section limits all obstructions to the second order.

Finally, we compute the $z$-spectral data, namely the scattering function $B(\zeta)$.
In this case, we can compute $B$ directly for a first-order perturbation, rather
than just deducing it up to an overall constant. We take the asymptotic gauge to be
\begin{equation}
   \Phi=2\pi\ii z\sigma_3 + O(z^{-2}), \quad 
           A_j=\pi\ii(y,-x,0)\sigma_3 + O(z^{-1})\quad\mbox{as $z\to\pm\infty$};
\end{equation}
and the basis solution vectors to be determined by
\begin{eqnarray*}
  \psi\,\exp(\pi z^2) &\to& \left[\begin{array}{c} 0 \\
         \exp(-\ii\pi\zeta y)\end{array}\right] \mbox{ as $z\to\infty$},\\
  \psi_{-}\,\exp(-\pi z^2) &=& \left[\begin{array}{c} \exp(\ii\pi\zeta y) \\
          0 \end{array}\right] \mbox{ for all $z\in\RR$}.
\end{eqnarray*}
Then we get, to first order in the perturbation,
\begin{equation}
 B(\zeta)=C_3[\vartheta_3(\pi\zeta)]^2+C_4[\vartheta_1(\pi\zeta)]^2,
\end{equation}
modulo an overall phase corresponding to the residual diagonal gauge freedom.
In this case, there are two spectral lines, which coincide if $C_3=0$ or $C_4=0$.
Note that the Higgs field $\Phi$ is constructed from exactly this combination
of theta-functions, and so again its zeros (and hence the monopoles) lie on the
spectral lines.


\subsection{Kuranishi Complex}
Here we revisit the perturbation calculation in a slightly more abstract setting,
adapting the similar instanton deformation argument of \cite{AHS78}.  In a more
general form, that is applicable to problems in general relativity and in gauge theory,
this argument appears in \cite{AMM}, and it is applied to the study of Yang-Mills
on a Riemann surface in \cite{Huebschmann}.

If both $(A,\Phi)$ and $(A+a, \Phi+\phi)$ satisfy the Bogomolny equation,
then
\begin{equation}\label{Eq:Def}
D_A a+*\big(D_A\phi-[\Phi, a]\big)+a\wedge a-*[\phi, a]=0.
\end{equation}
To simplify our notation let ${\bf A}=(A,\Phi)$ and ${\bf a}=(a,\phi)$, and
denote the linearized operator by $\delta_1 {\bf a}=D_A a+*(D_A \phi-[\Phi, a])$.
Also let $\{{\bf a}, {\bf a}\}=a\wedge a-*[\phi,a]$, so that Eq.~\eqref{Eq:Def} reads 
$\delta_1 {\bf a}+\{{\bf a}, {\bf a}\}=0.$

The space of linear deformations (up to gauge transformations) is given by
the middle cohomology of the complex
\begin{equation}\label{LinComp}
0\rightarrow\Lambda^0\xrightarrow{\delta_0} \Lambda^1\xrightarrow{\delta_1}
   \Lambda^2\rightarrow0,
\end{equation}
with $\delta_{0}:\Lambda\mapsto(D\Lambda,[\Phi,\Lambda]).$
We denote the space of its harmonic representatives by
\begin{equation}
{\rm Lin}=\{{\bf a}\in\Lambda^1 |\, \delta_1{\bf a}=0\ \text{and}\  \delta_0^*{\bf a}=0\}.
\end{equation}
Here the coclosure $\delta_0^*{\bf a}=0$ is a gauge fixing condition.

The total space of solutions, on the other hand, is 
\begin{equation}
{\rm Sol}=\{{\bf a}\in\Lambda^1 |\, \delta_1{\bf a}+\{{\bf a},
   {\bf a}\}=0\ \text{and}\  \delta_0^*{\bf a}=0\}.
\end{equation}
In the case of monopole walls, each of these spaces comes equipped with a
hyperk\"ahler structure.

Consider the covariant Laplacian of the complex \eqref{LinComp}, namely
$\Delta=\delta_1^*\delta_1+\delta_0\delta_0^*=D_A^2+[\Phi,[\Phi,\cdot]]$,
and let $G$ denote its Green's function, {\sl i.e.}\ $G\Delta=1-P$ with
$P$ a projection operator on the space of harmonic representatives.
Now consider the Kuranishi map
\begin{align}
F: {\bf a}\mapsto {\bf a}+\delta_1^*G\{{\bf a}, {\bf a}\}.
\end{align}
For an irreducible solution ${\bf A}=(A,\Phi)$, the map
$F: {\rm Sol}\rightarrow{\rm Lin}$ is one-to-one, and thus the linear analysis
provides the correct count of the moduli.  For a reducible solution ${\bf A}$, however,
the stabilizer group ${\rm Stab}_{\bf A}$ of the solution acts on Lin.  This action is triholomorphic.  
The only obstruction for extending the linear deformation is exactly the moment
map of this action, thus ${\rm Sol}$ is modeled by a hyperk\"ahler
reduction of ${\rm Lin}$ by the stabilizer group.  Thus the local geometry of the moduli space near the point ${\bf A}$
is ${\rm Lin}/\!\!/\!\!/{\rm Stab}_{\bf A}.$

In our explicit analysis of the previous subsection we find that the space ${\rm Lin}$ of linear deformations is eight-dimensional near the constant energy solution \eqref{const-energy}.  The stabilizer of this solution is $U(1),$ it is one-dimensional, thus, the linear analysis around this solution overcounts by one quaternionic dimension.  Therefore, this general theory gives the resulting dimension of the moduli space equal to four, in agreement with our explicit analysis.

\section{Conclusions}
In our study of doubly-periodic monopoles, we find that the Newton polygon provides
the most natural way of encoding their charges and singularity structure.  It also
delivers an immediate answer to the moduli counting problem: the number of $L^2$
moduli of a doubly-periodic monopole is four times the number of integer internal
points of the corresponding Newton polygon.

The asymptotic parameters, consisting of the subleading terms of the Higgs field
asymptotics, the asymptotic holonomy, and the singularity positions, correspond
to the perimeter points of the Newton polygon.  We give a number of illustrative
examples, and verify the Newton polygon count of the moduli for a particular
U(2) monopole wall.

Employing a string-theory picture, we identify any monopole wall with a D-brane
configuration.  The Coulomb branch of the gauge theory on this D-brane is
identified with the moduli space of the monopole wall.  The gauge-theory
computation of the asymptotic metric on such moduli spaces appeared
in \cite{arXiv:1107.2847}.

The next natural step would be to explore the dynamics of a monopole wall.
Following the argument in  \cite{NSF-ITP-81-116}, at low energies the dynamics
is given by the geodesic motion on its moduli space. The Newton polygon,
its amoeba, and its tropical degeneration appear useful in describing this
problem as well.

Another intriguing connection is the relation of doubly-periodic monopoles
to Calabi-Yau three-folds.  We map any monopole wall to a brane configuration
and a corresponding gauge theory.  The same theory can be obtained via
geometric engineering.  Under this correspondence, the moduli space we study
emerges as the moduli space of string theory on that Calabi-Yau space.

Any doubly-periodic monopole is defined over the  $T^2\times\mathbb{R}$ base space.
Considering $\mathbb{R}^3$ as a universal cover of the latter, we obtain a
BPS configuration in three-space with a constant magnetic field on one side, and
a possibly-different magnetic field on the other.  This is a monopole wall.
In this paper we did not distinguish these two, and we used the expressions
`doubly-periodic monopole' and `monopole wall' interchangeably.  While any
excitation of a doubly-periodic monopole can be viewed as an excitation of
a monopole wall, one expects the latter to have more excitations, which are
not necessarily doubly-periodic. It would be interesting to explore and compare
these two situations.

\bigskip\noindent{\bf Acknowledgments.}
RSW was supported by EPSRC grant EP/G038775/1 and STFC grant ST/J000426/1.
SCh was supported in part by the Simons Center for Geometry and Physics in Stony Brook.  
We benefited from the workshop ``Complex Geometry and Gauge Theory'' at Leeds University, July 2011, and LMS workshop on Geometry and Physics at Durham, September 2011.
SCh is grateful to Benoit Charbonneau, Samuel Grushevsky, Nigel Hitchin,
Jacques Hurtubise, Oleg Viro, and Edward Witten for illuminating  discussions.  

\appendix
\section{A four-dimensional detour}
One may view a doubly-periodic monopole as a limiting case of a self-dual
connection on a four-torus.  Namely, we consider a self-dual connection
$A$ on $T^2\times E_L$, where $T^2$ is the two-torus parameterized by
$x\sim x+1$ and $y\sim y+1$, and $E_L$ is a two-torus parameterized by
$z\sim z+L$ and $w\sim w+1/L$.  If this connection can locally be put in
a gauge such that it is $w$-independent, then the limiting connection as
$L\rightarrow\infty$ is a doubly-periodic monopole. The doubly-periodic
monopoles obtained in this way will have rather special properties as
$z\to\pm\infty$ (more on this below).

The Nahm transform of a self-dual connection $A$ on $T^2\times E_L$ is a
self-dual connection $A'$ on $\tilde{T}^2\times \tilde{E}_{1/L},$ with
the dual torus $\tilde{T}$ parameterized by $x'\sim x'+1$ and $y'\sim y'+1$,
and the dual torus $\tilde{E}_{1/L}$ parameterized by $w'\sim w'+1/L$ and
$z'\sim z'+L$. We should note at this point that it is the coordinate
$w'$ that is dual to $z$, and the coordinate $z'$ that is dual to $w$;
this is to be consistent with our other conventions.  

By choosing a complex structure in which $T^2$ and $E_L$ are elliptic
curves, we can define two spectral curves: 
\begin{align}
\Sigma_z&\subset T^2\times \tilde{E}_{1/L}& &\text{and}
           &\Sigma_{xy}&\subset \tilde{T}^2\times E_L,
\end{align}
given respectively by the eigenvalues of the monodromy along $E_L$
and $T^2$.  Similarly, we define spectral curves for the dual
connection $A'$: 
\begin{align}
\Sigma_z'&\subset \tilde{T}^2\times E_L& &\text{and} &\Sigma_{x'y'}
      &\subset T^2\times \tilde{E}_{1/L}.
\end{align}
These four curves are not independent, in fact $\Sigma_z=\Sigma_{x'y'}$
and $\Sigma_{z'}=\Sigma_{xy}$.
One might want to explore what  the remnants of $\Sigma_z$ and $\Sigma_{z'}$
are in the doubly-periodic monopole limit when $L\rightarrow\infty$.

Now we revisit $z$-scattering via this four-dimensional detour.
Let $T^2_{0}\subset T^2\times \tilde{E}_{1/L}$ denote the two-torus over
the point $z'=w'=0$ in $\tilde{E}_{1/L}$.  We denote the coordinates
of the points in $T^2_0\cap \Sigma_z$ by $(\xi_\rho(L),0)$ with
$\xi=x+\ii y$.  We expect that in the limit of infinite $L$ these
points of intersection of the spectral curve $\Sigma_z$ with the
$T^2$ fiber at zero tend to the spectral points $\xi_\rho$ defined
at the beginning of this section:
\begin{equation}
\lim_{L\rightarrow\infty}\xi_\rho(L)=\xi_\rho.
\end{equation}
Since $\Sigma_z=\Sigma_{x'y'}$ another way of finding these points
is given by the following prescription:
\begin{itemize}
\item look for a value $z'_\rho$ such that the dual Higgs field
$\Phi'(x',y',z'_\rho)$ has a vanishing eigenvalue for some value
of $x'$ and $y'$;
\item consider the eigenvalues of the monodromy of $A'$ on the
torus $\tilde{T}_{z'_\rho}$;
\item one of these eigenvalues corresponds to the zero
eigenspace of $\Phi'(x',y',z'_\rho)$ --- then this eigenvalue is
$\exp(2\pi i \xi_\rho)$, with $\xi_\rho$ the spectral point for
the $(A,\Phi)$ doubly-periodic monopole.
\end{itemize}
Surprisingly, it appears that the spectral points are related to the
zeros of the dual Higgs field.

\section{Theta-function Relations}
In our conventions $q=\exp(\ii\pi\tau)$ and the theta functions are as
in \cite{NIST}, so that 
\begin{align}
\theta_1(z,\tau)&=2\sum_{n=0}^\infty (-1)^n q^{(n+\frac{1}{2})^2}\sin((2n+1)z),&
\theta_3(z,\tau)&=1+2\sum_{n=1}^\infty (-1)^n q^{n^2}\cos(2nz).
\end{align}
They are related by 
\begin{equation}\label{Eq:13Shift}
\theta_1(z,\tau)=
  -\ii e^{\ii z}e^{\ii\pi\tau/4}\theta_3\big(z+\frac{\pi}{2}(1+\tau),\tau\big),
\end{equation} 
and are quasi-periodic:
\begin{align}
\theta_1\big(z+(m+n\tau)\pi,\tau\big)&=
   (-1)^{m+n}q^{-n^2}e^{-2\ii n z}\theta_1(z,\tau),\\
\theta_3\big(z+(m+n\tau)\pi,\tau\big)&=q^{-n^2}e^{-2\ii n z}\theta_3(z,\tau).
\end{align}
Under the modular transformation $\tau\mapsto\tau'=-1/\tau$ we have
\begin{align}\label{Eq:Mod}
\sqrt{-\ii\tau}\, \theta_1(z,\tau)&=
   -\ii e^{\ii\tau' z^2/\pi} \theta_1(z\tau',\tau),&
\sqrt{-\ii\tau}\, \theta_3(z,\tau)&= e^{\ii\tau' z^2/\pi} \theta_1(z\tau',\tau).
\end{align}

For a square period 1 torus parameterized by $\zeta=x+\ii y$, we define
$\vartheta_j(z)=\theta_j(z,\ii)$ for $j=1,3.$  We would like to demonstrate that
the following identity holds:
\begin{equation}\label{Eq:ThetaRelation}
\int_0^1\int_0^1e^{-4\pi y^2}\vartheta_i(\pi\zeta)^2
   \overline{\vartheta_j(\pi\zeta)}^2\,dx \,dy= \Upsilon \delta_{ij},
\end{equation}
where $\Upsilon$ is a constant and $\delta_{ij}$ is the Kronecker delta
with $i,j=1$ or $3.$

Applying the modular transformation $\tau'=-1/\tau, z'=\ii z$ sends $(x,y)$
to $(-y,x),$ giving
\begin{align}
&\mathop{\int}_{T^2} e^{-4\pi y^2}\vartheta_1(\pi\zeta)^2
  \overline{\vartheta_3(\pi\zeta)}^2\,dx\, dy=
\mathop{\int}_{T^2} e^{-4\pi x^2}\vartheta_1(-\ii \pi\zeta)^2
  \overline{\vartheta_3(-\ii \pi\zeta)}^2\,dx\, dy\\
\label{Eq:TranM}
&=\mathop{\int}_{T^2} (-1)e^{2\pi z^2}e^{2\pi \bar{z}^2}e^{-4\pi x^2}
  \vartheta_1(\pi\zeta)^2\overline{\vartheta_3(\pi\zeta)}^2\,dx\, dy=
  - \mathop{\int}_{T^2}e^{-4\pi y^2}\vartheta_1(\pi\zeta)^2
  \overline{\vartheta_3(\pi\zeta)}^2\,dx\, dy,
\end{align}
thus this integral vanishes.  Here in \eqref{Eq:TranM} we use the relation
\eqref{Eq:Mod} with $\tau=\ii$.

Next using \eqref{Eq:13Shift} and a change of variables
$\zeta'=\zeta+\frac{\pi}{2}(1+\ii)$, we have
\begin{align}
\Upsilon&=\mathop{\int}_{T^2} e^{-4\pi y^2}\vartheta_1(\pi\zeta)^2
 \overline{\vartheta_1(\pi\zeta)}^2\,dx\, dy=
\mathop{\int}_{T^2} e^{-4\pi y^2}e^{2\ii\pi(z+\bar{z}}e^{-\pi}
 \left|\vartheta_3(\pi\zeta+\frac{\pi}{2}(1+i))\right|^4\, dx\, dy\\
&=\mathop{\int}_{T^2} e^{-4\pi (y')^2}\vartheta_1(\pi\zeta')^2
 \overline{\vartheta_1(\pi\zeta)}^2\,dx'\, dy'.
\end{align}
This proves our main identity \eqref{Eq:ThetaRelation}.  Numerical evaluation
gives $\Upsilon=5.824747380908\ldots$.

\end{document}